\documentclass[final,3p,11pt]{elsarticle}

\usepackage{amssymb}
\usepackage{amsmath}
\usepackage{natbib}
\usepackage{lineno}
\usepackage{xcolor}
\usepackage{tabularray}

\usepackage{cite}
\usepackage[colorlinks=true,linkcolor=blue,allcolors=blue]{hyperref}%

\usepackage{epsfig,color,amsmath}
 
\usepackage{wrapfig}
\usepackage{setspace}  
\usepackage{xcolor}
\usepackage{epstopdf}
\usepackage{amssymb}
\usepackage{url}
\usepackage{mathtools}
\usepackage{enumitem}
\usepackage{multirow}
\usepackage{makecell}
\usepackage{amsmath}
\usepackage{stackrel}
\usepackage{booktabs}
\usepackage[flushleft]{threeparttable}
\usepackage{makecell}

\usepackage{subfig}
\usepackage{graphicx}

\usepackage[linesnumbered,lined,boxed,commentsnumbered,ruled,longend]{algorithm2e}


\setlist[itemize]{leftmargin=*}

\makeatother

\newcommand{\cs}{{\cal s}}

\newcommand{\m}{\boldsymbol}
\allowdisplaybreaks[4]
\newcommand{\mr}[1]{\mathrm{#1}}

\usepackage{amsthm}

\usepackage[usestackEOL]{stackengine}
\stackMath
\usepackage[framemethod=TikZ]{mdframed}
\usepackage{bm}
\mdfdefinestyle{MyFrame}{%
	linecolor=black,
	outerlinewidth=1.5pt,
	roundcorner=1.5pt,
	innerrightmargin=5pt,
	innerleftmargin=5pt,}

\usepackage{tikz}
\usetikzlibrary{matrix,positioning,decorations.pathreplacing}

\usepackage[export]{adjustbox}
\usepackage{lipsum}
\usepackage{mathtools}
\usepackage{cuted}
\usepackage{physics}

\journal{Journal of Hydrology}

\begin{document}

\begin{frontmatter}



\title{Global Optimization-Based Calibration Algorithm for a 2D Distributed Hydrologic-Hydrodynamic and Water Quality Model}


\affiliation[1]{organization={University of São Paulo, Department of Hydraulic Engineering and Sanitation, São Carlos School of Engineering},
            addressline={Av. Trab. São Carlense, 400 - Centro}, 
            city={São Carlos},
            postcode={13566-590}, 
            state={São Paulo},
            country={Brazil}}
            
\affiliation[2]{organization={The University of Texas at San Antonio, College of Engineering and Integrated Design, School of Civil \& Environmental Engineering and Construction Management},
            addressline={One UTSA Circle, BSE 1.310}, 
            city={San Antonio},
            postcode={78249}, 
            state={Texas},
            country={United States of America}}     
            
\author[1,2]{Marcus Nóbrega {Gomes Jr.}
        }
\author[2]{Marcio Hofheinz Giacomoni
        }
\author[1]{Fabricio Alonso Richmond Navarro
        }
\author[1]{Eduardo Mario Mendiondo
        }
\begin{abstract}
Hydrodynamic models with rain-on-the-grid capabilities are usually computationally expensive. This makes the use of automatic calibration algorithms hard to apply due to the large number of model runs. However, with the recent advances in parallel processing, computational resources, and increasing high-resolution climatologic and GIS data, high-resolution hydrodynamic models can be used for optimization-based calibration. This paper presents a global optimization-based algorithm to calibrate a fully distributed hydrologic-hydrodynamic and water quality model (HydroPol2D) using observed data (i.e., discharge, or pollutant concentration) as input. The algorithm can find a near-optimal set of parameters to explain observed gauged data. The modeling framework presented here, although applied in a poorly-gauged catchment, can be adapted for catchments with more detailed observations. We applied the algorithm in different cases of the V-Tilted Catchment, the Wooden-Board catchment, and in an existing urban catchment with heterogeneous data. The results of automatic calibration indicate NSE = 0.99 for the V-Tilted catchment, RMSE = 830 $\mathrm{mgL^{-1}}$ for salt concentration pollutographs (i.e., 8.3\% of the event mean concentration), and NSE = 0.89 for the  urban catchment case study. This paper also explores the issue of equifinality in modeling calibration (EqMC). Equifinality is defined as the set of different parameter combinations that can provide equally good or accepted results, within the physical parameter ranges. EqMC decreases with the number of events and increases with the choice of partially or nonproducing runoff ones. Furthermore, results indicate that providing more accurate parameter ranges based on a priori knowledge of the catchment is fundamental to reduce the chances of finding a set of parameters with equifinality.


\end{abstract}

 \begin{highlights}
 \item An automatic optimization calibration algorithm for spatially-distributed flood and water quality is developed
 \item The algorithm uses HydroPol2D model and is able to calibrate water quantity and quality parameters globally
 \item Source data in observed gauges such as discharges, depths, and concentration is required for the calibration.
 \item Equifinality is investigated and increases with the use of non-producing runoff events and increases with poor gauge locations.
 \end{highlights}



\end{frontmatter}



\section{Introduction}
The advances in computational processing, high resolution GIS data availability, and relatively more complete physically-based models enables the application of fully distributed hydrodynamic and pollutant transport and fate models \citep{yang2010geospatial,gomes2022flood}. Although the application of fully distributed models (i.e., models that discretize the catchment domain into finite cells) remotes to the late 1970s \citep{zhang1990hydrologic}, the demands for high-resolution modeling, especially for flood and pollution assessment, make optimization-based calibration (i.e., herein referred to as automatic calibration) complex, time-consuming, and dependent on prior knowledge of the modeler \citep{blasone2008uncertainty} due to the relatively high computational effort required to perform numerical hydrodynamic simulations with high-resolution data \citep{blasone2008uncertainty,brath2004analysis}.

The complexity comes because the Shallow-Water-Equations dynamic problem forms a set of hyperbolic partial differential equations with no analytical solution for complex real-world cases \citep{bermudez1994upwind}, requiring finite-volume or finite-difference numerical schemes to solve the problem \citep{brunner2016hec}. The hydrodynamic problem can be simplified into diffusive-like problems when local acceleration and inertial terms can be neglected \citep{akan2021open} and these can reduce computational effort \citep{gomes2023hydropol2d}; however, the number of required simulations for a full calibration still makes the process laborious. Since the governing equations of flow and pollutant routing are generally performed for each element in the grid domain either as a matrix-wise expression or as an element-wise approach, the numerical modeling process can be challenging for finer mesh grids, such as the ones required for flood mapping \citep{do2023generalizing}. This might be one of the reasons why only a few articles attempted to develop automatic calibration algorithms for 2D hydrodynamic and pollutant transport and fate models \citep{afshar2011particle}.

The parameter discretization of distributed models is usually performed by the discrete categorical values of Land Use and Land Cover (LULC) and Soil type classifications. This information is entered as georeferenced maps, and each cell of the computational domain is assigned with the parameters associated with each input map. For example, in HydroPol2D \citep{gomes2023hydropol2d}, hydrological parameters (i.e. Green-Ampt parameters) are assigned with the Soil raster, and hydrodynamic (i.e., Manning's roughness coefficient and initial abstraction) and water quality (i.e., build-up and wash-off parameters) are assigned according to the LULC raster. 

The automatic calibration of distributed models can be a taunting task due to the degrees of freedom of the optimization problem and the number of calibration variables that could increase proportionally to the number of land uses and soil classifications \citep{debele2008coupling}. Moreover, due to the nonlinear behavior of hydrology and hydrodynamics, the use of convex optimization to find global optima is unfeasible, unless several simplifications are performed in the modeling equations \citep{wang2020new}. Additionally, defining the appropriate ranges for model parameters can also lead to unrealistic parameter estimations, especially when the physical boundaries of the parameters are incorrectly treated \citep{domeneghetti2012assessing}. Due to the complexity of automatic parameter estimations, several studies have successfully performed manual calibrations using distributed models \citep{ardiccliouglu2019calibration,phillips20051d,li2021high}. Despite these challenges, calibrating a complex model with a relatively large number of parameters can be even more complex.

Although successful calibrations are presented in the literature, one of the yet unsolved and considerably complex problem is the total reduction of equifinality for a relatively large number of model parameters and ranges \citep{fatichi2016overview}. This paper does not attempt to provide a definitive solution to this issue; rather, we explore the factors associated with the chances of finding parameter equifinality in automatic calibration. Investigating the trade-offs between the number of observation points, intensity of rainfall events, and combination of different associations of gauges that can affect parameter equifinality.

Several models physically-based models are available in the literature, such as the Hydrologic Engineering Center - River System Analysis \citep{brunner2016hec}, the Stormwater Management Model - SWMM \citep{rossman2010storm}, and the Gridded Surface and Subsurface Analysis \citep{downer2004gssha}. However, only a few studies used the aforementioned models or developed new models that can take advantage of automatic calibration capabilities \citep{cho2015watershed,dung2011multi,hong2019physically}. 

Research conducted in \citep{cho2015watershed} used a genetic algorithm solver to calibrate observations with modeling results; however, as most studies in automatic calibration of hydrologic models \citep{gupta1999status, confesor2007automatic} they used a semi-distributed model that cannot account for some important hydrodynamic features such as backwater effects or hydrologic characteristics such as spatial distribution of soil moisture and pollutants inside the subcatchments. Other recent research using the SWMM model attempts to develop automatic calibration algorithms for semi-distributed models, as shown in \citep{behrouz2020new,swathi2019automatic}. 

The research conducted in \citep{hong2019physically}, however, considered a physically based fully distributed model that assumes various wash-off processes such as detachment and transport of particulate, resulting in six wash-off parameters that, in addition to the water quantity model parameters, must be calibrated altogether. This dramatically increases the decision variable space and might result in relatively longer simulations, as well as increase the chance of finding a different set of parameters that could explain the modeling results within a defined physically-based parameter range (i.e., Equifinality effect \citep{beven2001equifinality}). In addition to calibrating the model, an essential part is the model validation/evaluation that can be done to understand the model capability to represent the behavior of the system outside the calibration range \citep{shen2022time}.

The validation of conceptual lumped-paramater hydrological models as the ones presented in \citep{shen2022time} does not require detailed description of the surface topography. For fully distributed models, however, if all parameters are correct but the Digital Elevation Model (DEM) does not allow proper continuity of the flow, the model performance is affected. Poor DEMs are one of the limiting factors of applyng distributed models. DEMs usually contain noises, bridges, and are affected by vegetation \citep{hawker2018perspectives}. Raster-based algorithms such as HydroPol2D might be hydraulically affected by such issues in the DEM, requiring a pre-processing filtering to allow proper flow paths and continuity. To this end, filtering algorithms to smooth DEM flow paths \citep{schwanghart2014topotoolbox}, remove vegetation noises \citep{de2013large}, reduce sharp elevations \citep{conrad2015system}, or smooth hillslopes \citep{milledge2009potential}, can be applied to treat DEMs.

\subsection{Paper objectives and Contributions}
As shown above in the literature, although several studies successfully calibrated distributed hydrodynamic models, most of the calibration studies were performed manually. With advances in computational processing and parallel computing, models that take advantage of these techniques can be applied and used for automatic optimization-based calibration. Only a few studies developed automatic calibration algorithms for fully distributed, high-resolution, hydrological-hydrodynamic models. This is likely due to complexity of due to the complex computational models with a high number of cells, states, and relatively high nonlinear underlying physical laws. 

The objective of this paper is to derive a flexible framework to apply a formal HydroPol2D (Hydrodynamic and Pollution 2D model) calibration-optimization problem (i.e., a 2D distributed water quantity and quality model) using only source data at observed gauges as input. Although we use HydroPol2D in this paper, the methods developed here are valid for any other hydrodynamic model used to estimate information at gauge stations.

More specifically, in this paper, we develop a modeling framework that calibrates HydroPol2D for gauged information such as hydrographs and / or stage hydrographs and / or pollutographs using rainfall, initial soil moisture and initial water surface depths from the warm-up process as initial conditions for the model. All other hydrological-hydrodynamic and pollutant transport and fate parameters can be automatically obtained by the calibrator module developed in this paper. The method is of particular interest in catchments that already have point source gauged data in observed nodes (e.g., outlet), and these data can be used to generate spatiotemporal information within the catchment by running a calibrated HydroPol2D model in the catchment \citep{brath2004analysis}. 

The fundamental contributions of this paper are:
\begin{itemize}
    \item We develop an automatic calibration routine to estimate HydroPol2D parameters requiring only point-source information such as depths, flows, or pollutant concentrations. 
    \item We provide a framework capable of calibrating HydroPol2D for various events for water quantity and/or water quality modeling.
    \item We improved the model presented in \citep{gomes2023hydropol2d} by allowing not only a Von-Neuman (4-D) cell topology but also by adding a Moore grid (8-D) topology (i.e., four in
the cardinal direction and four in the diagonal direction), allowing cells to have more flow directions, eventually decreasing the flow paths.
    \item We also expand \citep{gomes2023hydropol2d} by allowing raster-based DEM pre-processing algorithms to enhance flow continuity in coarse resolution DEMs.
\end{itemize}

\section{Model Background}
HydroPol2D model is a 2D model of transport and fate of hydrodynamic and water quality. The watershed is discretized into finite cells with known resolution $\Delta x$ and the equations of conservation of runoff mass, momentum and conservation and transport of pollutant mass are all solved matrix-wise. For a more complete description of the model, please refer to \citep{gomes2023hydropol2d}. Let $n$ and $p$ be the number of rows and columns in the watershed domain. In this paper, we show the main equations for the 2D surface water dynamics in Eq.~\eqref{equ:explicit_scheme_overland} and 2D pollutant transport \eqref{equ:buildup}-\eqref{equ:washoff_finite}. To ensure proper mathematical notation for the modeling equations, we introduce the paper notation as follows:

\vspace{0.5cm}
\noindent \textit{\textbf{Paper's Notation:}} 
Italicized, boldface upper and lower case characters in boldface represent matrices and column vectors: $a$ is a scalar, $\m a$ is a vector, and $\m A$ is a matrix. Matrix $\m I_n$ denotes an identity square matrix of dimension $n$-by-$n$, while $\m O_{m \times n}$ and $\m 1_{m \times n}$ denotes a zero and one matrix of size $m$-by-$n$, respectively.
The notations $\mathbb{R}$ and $\mathbb{R}_{++}$ denote the set of real and positive real numbers. The notations $\mathbb{R}^n$ and $\mathbb{R}^{m\times n}$ denote a column vector with $n$ elements and an $m$-by-$n$ matrix in $\mathbb{R}$. The element-wise product or Hadamard product is defined as $\m x \circ \m y \coloneqq [x_1 y_1, x_2 y_2, \dots{}, x_n y_n]^T$ multiplications. Similarly, the element-wise division or Hadamard is defined as $\m x \oslash \m y \coloneqq [ \frac{x_1}{y_1},\frac{x_2}{y_n},\dots{},\frac{x_n}{y_n}]^T $.
The element-wise $p$ power of a matrix $\m A, $ $(\m A^{\circ p})$, with $\m A \in \mathbb{R}^{m \times n}$ and $p \in \mathbb{R}$  is given by $ a_{i,j}^p$ for $i \in \mathbb{N}_{++}$, and $ j \in \mathbb{N}_{++}$.

\begin{align} \label{equ:explicit_scheme_overland}
    \m H(t + \Delta t) = &\overbrace{\m H(t) + \Delta t \left(\m B_{\mr i} \m I(t) + \frac{1}{A}\m B_{\mr Q} \m Q(t) - \m F(\m H(t), \m F_{\mr d}(t), \m E_{\mr{TR}}) - \m E_{\mr{TR}}(t)\right)}^{\m H_{\mr{ef}}(t)} 
 \notag \\ &+ \Delta t \sum_{i=1}^{m}\left[\m Q_{\mr{in}}^i\Bigl(\m H(t)\Bigr) - \m Q_{\mr{out}}^i\Bigl(\m H(t)\Bigr)\right],
\end{align}
where $\m H_{\mathrm{ef}}(t) \in \mathbb{R}^{\mr{n \times p}}$ is the effective depth for overland flow routing, $\m B_i$ defines the cells that receive rain-on-the-grid boundary condition values from $\m I$, $\m B_{\mathrm{Q}}$ defines the cells that receive inflow hydrograph boundary conditions from $\m Q$, $\m F$ is the infiltration rate, $\m F_d$ is the cumulative infiltration depth, $\m E_{\mr{TR}}$ is the evapotranspiration rate, $\m Q^i_{\mr{in}}$ is the inflow rate for direction $i$, calculated from the outflow rate $\m Q_{\mr{out}}$ with the time-varying flow-direction matrix.

To solve Eq.~\eqref{equ:explicit_scheme_overland}, we develop a weighted cellular automata approach using Manning's equation to estimate matrix $\m Q_{\mr{out}} \in \mathbb{R}^{\mr{n \times p \times m}}$, with $m$ being the number of flow directions (i.e., 4 or 8), and using topological relationships between cells, we derive $\m Q_{in}(t) \in \mathbb{R}^{\mr{n \times p \times m}}$ in terms of $\m Q_{\mr{out}}(t) \in \mathbb{R}^{\mr{n \times p}}$ by calculating the sparse time-varying flow direction matrix $\m B_d(\m H(t)) \in \mathbb{R}^{\mr{np \times np \times m}}$. $\m F(\m H(t),\m F_d(t),\m E_{\mr{TR}}) \in \mathbb{R}^{\mr{n \times p}}$ is the infiltration rate, which is calculated with the Green-Ampt model \citep{green1911studies} and depends on soil hydraulic properties. Details of how to solve the problem can be found \citep{gomes2023hydropol2d,guidolin2016weighted} and a pseudo-code of the Celular-Automata (CA) used in this paper is presented in Algorithm \ref{alg:automata}. In brief, the CA algorithm calculates the friction slope to the steepest water surface direction using Manning's equation, determines the maximum velocity to this direction, uses a weighted-average system based on the available void volume of neighbor cells, and distributes the runoff according to the weights given for each neighbor. In addition, it restricts the maximum outflow volume to avoid checkerboard oscillations between two adjacent cells \citep{hunter2005adaptive}.

\begin{algorithm} 
\normalsize
	\caption{Celular automata pseudocode} \label{alg:automata}
	        \textbf{input:} Cell elevations $(\m E)$, initial surface water depths $(\m{ \mr{WSE}})$, matrices of Manning's roughness $(\m N)$, initial abstraction $(\m H_0)$, time-step $(\Delta t)$, grid discretization $(\Delta x)$, friction slope at the outelt for normal flow $(s_0^b)$, large number $(c)$, Velocity to the steepest direction $ (\m V_{\mathrm{m}})$, Intercell Volume $(\m I_{tot})$ outflow volumes from previous time-step $(\m I_{tot}^p)$, Minimum water depth $(\Delta h_{\mr{min}})$ Set of cells in the catchment domain $(\mathbb{C})$, set of cells at the outlet $(\mathcal{O})$, Domain borders $(\mathbb{B})$, number of flow directions $(m)$, gravity acceleration ($g$), cell area ($A$). \\
            \For{i = 1 to m}{
            \textbf{compute: }$\Delta \m H_{\mr{ef,i}} =  \text{\textbf{WSE}} - \text{\textbf{WSE}}_i$, $ \Delta \m H_{\mr{ef}} \in \mathbb{R}^{\mr{n \times p \times (m+1)}}, \textbf{WSE} \in \mathbb{R}^{\mr{n \times p}}$\\
            }
	        \eIf {Outlet Type = 1}{
	        \textbf{compute: }$\Delta \m H_{\mr{ef,m+1}} = s_0^b \Delta x$ $\forall~\mathbb{C} \in \mathbb{O}$    
	        }{
	        \textbf{compute: }$\Delta \m H_{\mr{ef,m+1}} = \m H_{\mr{ef}}^{\circ -1/6}g^{0.5}\circ \m N$ $\forall~\mathbb{C} \in \mathbb{O}$} 
	        $\m H_{\mr{ef,m+1}} \leftarrow 0~\forall~\mathbb{C} \in \mathbb{B}$ \\
	        $\m \Delta \m H_{\mr{ef}} \leftarrow 0~\forall~\Delta \m H_{\mr{ef}} \leq \Delta h_{\mr{min}}$\\
	        \textbf{compute: } $\Delta \m V = A\Delta \m H_{\mr{ef}},~\Delta \m V \in \mathbb{R}^{\mr{n \times p \times (m+1)}} $\\
	        $\Delta \m V \leftarrow c,~\forall~\Delta \m V = 0$\\
	        \textbf{compute: } $\Delta \m V_{\mr{max}} = \max{(\Delta \m V)},~\Delta \m V_{\mr{max}} \in \mathbb{R}^{\mr{n \times p}}$ \\
	        \textbf{compute: } $\Delta \m H_{\mr{ef,max}} = \max{(\Delta \m H_{\mr{ef}})},~\Delta \m H_{\mr{ef,max}} \in \mathbb{R}^{\mr{n \times p}}$ \\
	        \textbf{compute: } $\Delta \m V_{\mr{min}} = \min{(\Delta \m V)},~\Delta \m V_{\mr{min}} \in \mathbb{R}^{\mr{n \times p}}$ \\
	        \textbf{compute: }$\m \Omega = (\Delta \m V_{\mr{tot}} + \Delta \m V_{\mr{min}})\oslash \Delta \m V,~\m \Omega \in \mathbb{R}^{\mr{n \times p \times (m+1)}}$ \\ 
	        \textbf{compute: } $\m \Omega_{\mr{max}} = \max{(\m \Omega)},~\m \Omega_{\mr{max}} \in \mathbb{R}^{\mr{n \times p}}$ \\
	        \textbf{compute: } $\m V_{\mathrm{m}} = \min{(\sqrt{g}\m H_{\mr{ef}}^{\circ 0.5}, \m N \oslash \max{(\m H_{\mr{ef}} - \m H_0)^{\circ 2/3}\circ (\m H_{\mr{ef,max}}(1/\Delta x)})^{\circ 0.5})},~\m V_{\mathrm{m}} \in \mathbb{R}^{\mr{n \times p}}$ \label{equ:manning}\\
	        \textbf{compute: } $\m I_{\mr{tot}}^* = \min{(\omega \m H_{\mr{ef}}, (\Delta x / \Delta t) \m V_{\mathrm{m}} \circ \m H_{\mr{ef}}, \m I_{\mr{tot}}^{p} + \Delta \m V_{\mr{min}})},~\m I_{\mr{tot}}^* \in \mathbb{R}^{\mr{n \times p}}$\\
	        \textbf{compute: } $\m I_{\mr{tot}}^* \leftarrow \text{sum}_{3} ( \m \Omega \circ \m I_{\mr{tot}}^*$)\\
	        \textbf{compute: } $\m Q_{\mr{out}} = 1/(\Delta t A)\m I_{\mr{tot}}^*,~\m Q_{\mr{out}} \in \mathbb{R}^{\mr{n \times p \times m}}$ \\
	        \textbf{compute: } $\m H_{\mr{ef}} \leftarrow \m H_{\mr{ef}} - (1/\omega) \m I_{\mr{tot}}^*$ \\
	        \textbf{output: } $\m Q_{\mr{out}}~\mathrm{from~Eq.~\eqref{equ:explicit_scheme_overland}},~,\m H_{\mr{ef}},~ \m I_{\mr{tot}}^*$
\end{algorithm}

During dry weather periods, the initial mass of pollutant available in the catchment domain varies according to each Land Use and Land Cover (LULC) classification \citep{rossman2010storm}, and can be calculated using the build-up equation as follows:

\begin{equation} \label{equ:buildup}
    \m B(t + \text{ADD}) = \m C_1 \circ (\m 1_{\mr{n \times p}} - \m e_{\mr{n \times p}}^{\circ \m C_3 \text{ADD}}) + \m B(t),
\end{equation}
where $\m C_1 \in \mathbb{R}^{\mr{n \times p}}$ is the buildup maximum concentration (kg.ha\textsuperscript{-1}), $\m C_2 \in \mathbb{R}^{\mr{n \times p}}$ is a fitted decreasing factor (day\textsuperscript{-1}) and ADD is the antecedent dry days prior to the event (days).

During wet-weather periods, the pollutant mass balance equation is calculated in terms of the discharges in each direction and is given by:  

 \begin{equation} \label{equ:washoff_finite}
     \sum_{l=1}^{m}{\m B_{\mr{out}}^i}(t) = \m W_{\mr{out}}^{\mr{tot}}(t ) = \Delta t \m \sum_{i=1}^{m}\left [\m C_3 \circ \m (\m Q_{\mr{out}}^i(\m H(t))^{\circ \m C_4} \circ \m B(t)\right],
 \end{equation}
 where $\m W_{\mr{out}}^{\mr{tot}}(t) \in \mathbb{R}^{\mr{\mathrm{n \times p}}}$ is the total amount of pollutant in kg that left each cell considering all directions.

 The available pollutant mass is calculated using Euler's forward finite difference scheme applied in Eq.~\eqref{equ:washoff_finite}, such that:

 \begin{equation} \label{equ:pollutant_finite}
    \m B(t+\Delta t) = \m B(t) + \Bigl[\m B_d(t) \m W_{out}(t)\Bigr],
 \end{equation}
 where $\m W_{\mr{out}}(t) \in \mathbb{R}^{\mr{n \times p \times m}}$ is the pollutant washed mass for each Von-Neumann or Moore  \citep{torres2022conceptual} direction (i.e., moving from 1 to $m$ clockwise, with 1 being the leftwards direction).

\subsection{Decision Variables in HydroPol2D Automatic Calibration Problem}
In HydroPol2D, parameters are spatially derived as a function of the Land Use and Land Cover (LULC) and Soil rasters. Minimum and maximum values of each decision variable are entered for each classification of these rasters. Let $n_l$ be the number of land use classifications and $n_s$ be the number of soil classifications in a catchment. Also, let $n_v^l$ be the number of decision variables related to the LULC map and $n_v^s$ be those related to the soil map; therefore, we have a decision vector $\m x \in \mathbb{R}^{n_x}$ with $n_x = n_l n_v^l + n_s n_v^s.$

We classify the decision variables into water quantity variables (superscript $r$), water quality variables (superscript $w$), and soil related parameters (subscript $s$). Furthermore, we classify the variables as LULC-based and soil-based into subscripts 1 to $n_l$ and 1 to $n_s$, respectively. The decision variable of the calibration problem can be written as $\m x = [\m x_l^q,~\m x_l^w,~\m x_s^q]^{\mr{T}}$, such that:
\begin{subequations}
    \begin{align}
        \m x_l^r &= [n_1,~\dots,n_{n_l},~h_{0,1},~\dots,~h_{0,n_l}]^{\mr T} \\ 
        \m x_l^w &= [C_{1,1},~\dots,C_{1,n_l},~C_{2,1},~\dots,~C_{2,n_l}, C_{3,1}, \dots C_{3,n_l},~C_{4,1},~\dots C_{4,n_l}]^{\mr T} \\  
        \m x_s &= [k_{sat,1},~\dots,k_{sat,n_s},~\Delta \theta_{1},~\dots,~\Delta \theta_{n_s},\psi_1,~\dots \psi_{n_s}]^{\mr T},
    \end{align}
\end{subequations}
where $n$ is the Manning's roughness coefficient ($\mr{sm^{-1/3}}$), $h_0$ is the initial abstraction (mm), $k_{\mr{sat}}$ is the saturated hydraulic conductivity of the soil ($\mr{mmh^{-1}}$), $\Delta \theta$ is the soil moisture deficit (-), and $\psi$ is the soil suction head (mm), $C_1$ is the build-up coefficient, $C_2$ is the build-up.

\subsection{Initial Values for the Simulation}
In addition to the parameters, in the automatic calibrator of the HydroPol2D model, we can enter initial maps of water surface depth and soil moisture to accurately represent the initial conditions for simulation, for each event. The code is designed to calibrate at last 10 events with 10 observation points of discharge, water depth, and pollutant concentration. Another input map that could be entered is the distributed pollutant mass in the catchment domain prior to the simulation; however, in this paper we opted to calculate this mass in terms of the antecedent dry days and assume that it is uniformly distributed according to the land use classification.

\subsection{Fitness Functions}
In this section we show the fitness functions allowed in the automatic calibrator. These functions can be used for the calibration of hydrographs, stage-hydrographs, and pollutographs at the catchment outlet. Detailed mathematical descriptions of these functions are available in the Supplemental Material. In this paper, we use the Nash-Sutcliffe-Efficiency (NSE) \citep{nash1970river}, the coefficient of determination ($r^2$),  the root-mean-square-error (RMSE) \citep{fisher1920012}, the Peak Flow Bias, and the relative runoff volume error.

In addition to the previously defined functions, users can have the flexibility to write their own fitness function since all codes are open source. Some examples that are predefined in the model and not fully detailed here for the sake of parsimony are the (i) mean average error, (ii) event mean concentration, (iii) PBIAS, and (iv) runoff volume mean error.

\subsection{Optimization Constraints}
The optimization problem is subject to 4 constraints. First, the model is constrained to HydroPol2D dynamical model that has conservation of mass and conservation of momentum constraints and pollutant transport and fate dynamics. Moreover, the optimization problem is constrained to equality constraints (e.g., the case where a parameter is known or cannot vary). In addition, we can add equality constraints in the model and, finally, we can add the minimum and maximum parameter ranges for each decision variable.

\subsection{Objective Function}
We can calibrate a single or multiple events together. Therefore, let the index $j$ represent the j-th event used for calibration. Moreover, let $f$ collect fitness functions such that a possible scenario of objective functions could be $f_1 = \mr{-NSE}$, $f_2 = -r^2$, $f_3 = \mr{-RMSE}$, and $f_4 = \mr{\eta_q}$, for example. The problem is set as a single-objective minimization problem, and the objective function used in this paper can be written as a function of linear combinations between each individual fitness function for each event, such that:

\begin{equation} \label{equ:objective_function}
    O_f =  \sum_{j=1}^{n_{e}} \beta_j \Bigl[ \Bigr (\sum_{i = 1}^{n_f}\alpha_{i} f_i  \Bigr )\Bigr ],
\end{equation}
where $n_f$ is the number of fitness functions used in the optimization $\alpha$ defines the relative weight of each individual function $f_i$, $n_e$ is the number of events used in the calibration, and $B_j$ is the weight given by the objective function values for each event.

It is important to mention that the factor $\alpha_i$ must be such that it normalizes the varied objective functions used, to avoid over-weighting in fitness functions with different scales of magnitude and units.  

\subsection{Automatic Calibration Optimization Problem}
In this section, we define the automatic calibration optimization problem. Although the nature of hydrological model calibration can be inherently multiobjective \citep{shafii2009multi}, for the sake of parsimony and to allow practical application, we focus on developing a single objective automatic calibration problem. 
It can be written by minimizing the objective function, subject to HydroPol2D dynamics. Users can define equality constraints in case few of the parameters are known or inequality constraints whenever a linear relationship among them is available, such that we can write the problem as:

\begin{subequations} \label{equ:optimization_problem}
\begin{align}
\min_{\m x} ~ &  O_f =  \sum_{j=1}^{n_{e}} \beta_j \Bigl[ \Bigr (\sum_{i = 1}^{n_f}\alpha_{i} f_i  \Bigr )\Bigr ]\\
&  \label{equ:objective_functions}\\
\textrm{s.t.}~~ &\mathrm{HydroPol2D~Dynamics}~\mr{in}~\mr{Eqs.}~\eqref{equ:explicit_scheme_overland}-\eqref{equ:pollutant_finite}
 \\
&\m A_{\mathrm{eq}}\m x = \m B_{\mathrm{eq}} \label{equ:equality_constraint}\\ 
& \m x_l \leq \m x \leq \m x_m
\end{align}
\end{subequations}

The problem posed in Eq.~\eqref{equ:optimization_problem} is non-linear and non-convex. The use of evolutionary strategies such as the Shuffled Complex Evolution (SCE) \citep{naeini2019three} has been used for this type of calibration problems \citep{tigkas2016comparative}. In Matlab, several solvers are available to solve problems as Eq.~\eqref{equ:optimization_problem}, such as \textit{Global-Search}, \textit{Patter-Search} or \textit{Genetic-Algorithms} (GA) \citep{higham2016matlab}. Herein, we choose GA due to its flexibility to deal with non-linear problems \citep{giacomoni2017multi} and its ability to use parallelization in Matlab. The GA is a population-based probabilistic optimization method that emulates the principles of genetics and natural selection \citep{tigkas2016comparative}. Since the goal of hydrologic-hydrodynamic model calibration is not essentially finding the global optima but a physically possible set of parameters trying to avoid equifinality, we set all problems to run with a relatively low number of generations, but still with a relatively large number of population to allow a proper exploration of the decision variable space. 

\subsubsection{Genetic Algorithm Properties}
\begin{figure*}
    \centering
    \includegraphics[scale = 0.3]{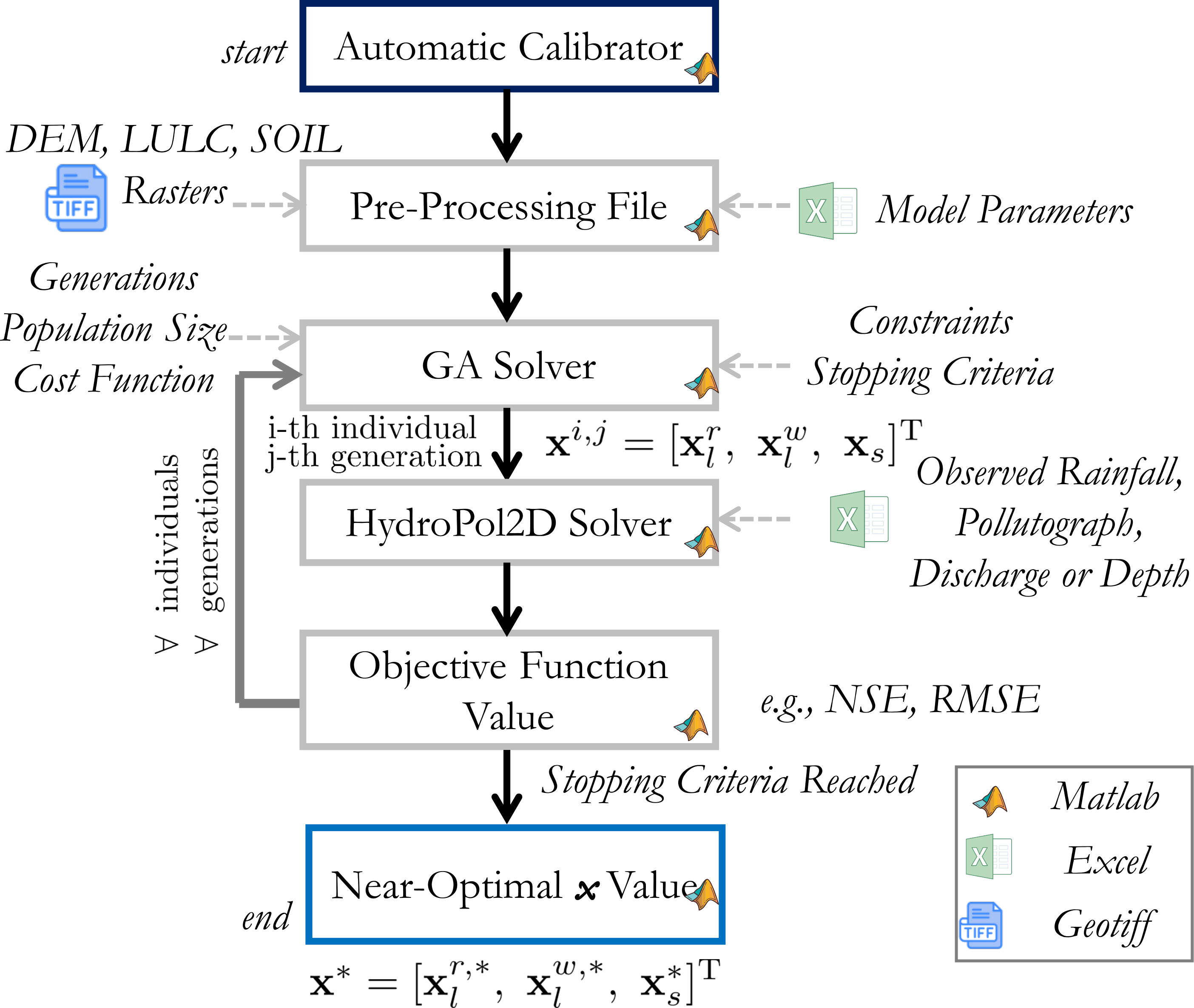}
    \caption{Automatic calibrator flowchart. First, the model reads the automatic calibration inputs, then it runs a pre-processing files, defining the required numerical input for HydroPol2D, such as the DEM, LULC, and Soil Maps. Following this phase, the model runs the GA solver, that uses HydroPol2D to estimate the objective function values and this process is looped until the stopping criteria is reached.}
    \label{fig:automatic_figure}
\end{figure*}

We set the problem to run for (10-40) generations with a 100 population. The stopping criteria were twofold: (i) simulation would stop if the number of maximum generations is reached, (ii) if no improvement in the objective function were found in 30 min. All parameters of the genetic algorithm are set as standard values from Matlab \citep{higham2016matlab}. A flowchart of the optimization process is presented in Fig.~\ref{fig:automatic_figure}.

\section{Case Studies}

\begin{figure*}
    \centering
    \includegraphics[scale = 0.27]{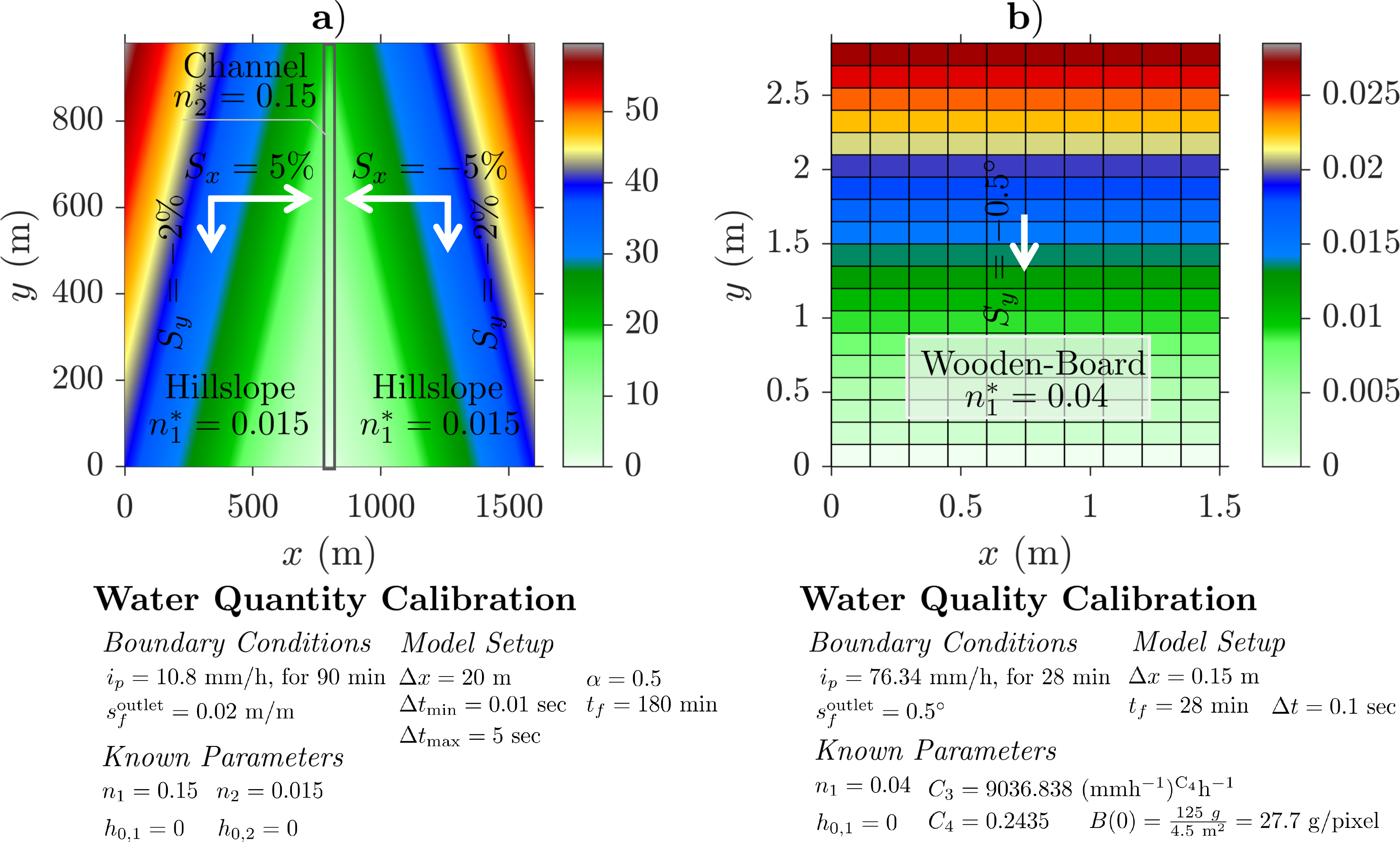}
    \caption{Numerical Case Study 1 and 2 digital elevation models in meters. Part a) is the V-Tilted Catchment whereas part b) is the Wooden-Board catchment. The parameters of each case are shown in the figure, where $i_p$ is the constant rainfall rate, $s_f^{\mr{outlet}}$ is the friction slope at the outlet, $\Delta x$ is the pixel size, $\Delta t$ is the constant time-step assumed, and $t_f$ is the end of the simulation.}
    \label{fig:case_study}
\end{figure*}

\subsection{Numerical Case Study 1 - V-Tilted Catchment}
The objective of this numerical case study is to test the ability of the optimization model to predict the Manning's roughness coefficient of the catchment and to check if the model can predict that there are no initial abstractions and no infiltration in this case study. Essentially, we want to answer the following question:

\begin{itemize}
    \item Q1: \textit{Does the automatic calibration algorithm can identify the Manning's roughness coefficients of hillslopes and main channel, as well as the initial abstraction values of these land uses? In addition, can it identify whether infiltration is being considered or not in this case study? }
\end{itemize}

We choose the V-Tilted catchment as a virtual experiment inverse problem \citep{fatichi2016overview}. The catchment has 4,050 cells of 20 x 20 m in size and has a reasonable fast computation, allowing the use of meta-euristics that rely on multiple computations of the objective function. To answer Q1, we define the decision vector of this problem as:

$$\m x = [n_1, n_2, h_{0,1}, h_{0,2}, k_{\mr{sat},1}, \Delta \theta_1, \psi_1]^{\mr T},$$ where subscripts 1 and 2 are the LULC classifications in the catchment (i.e., 1 are the hillslopes and 2 is the channel, see Fig.~\ref{fig:case_study}a)). It is assumed that there is only one type of soil in the catchment such that $n_s = 1$. To set only the water quantity variables as decision variables, we add an equality constraint in Eq.~\eqref{equ:equality_vtilted} by entering the known parameters in $\m B_{\mr{eq}}$. In this problem, we choose the NSE as the objective function, since we are focused on calibrating the modeled flow discharge with the observed discharge at the outlet. This case study has no infiltration or initial abstraction and is a reverse problem since we know the right parameters \citep{kollet2006integrated}; however, we decided to include infiltration variables in the optimization problem formulation to see if the algorithm can identify this condition. The formal optimization problem can be written as:

\begin{subequations} \label{equ:optimization_vtilted}
\begin{align}
\min_{\m x} ~     O_f =  & ~\mr{NSE}\\
&  \\
\textrm{s.t.}~~ &\mathrm{HydroPol2D~Dynamics}~\mr{in}~\mr{Eqs.}~\eqref{equ:explicit_scheme_overland}-\eqref{equ:pollutant_finite}
 \\
&\begin{bmatrix} \label{equ:equality_vtilted}
    \m O_{4} & \m O_{4\times 4} & \m O_{4 \times 3} \\
    \m O_{8\times 4} & \m I_{8} & \m O_{8 \times 3} \\
    \m O_{3 \times 4} & \m O_{3 \times 4 } & \m O_{3 \times 3}
\end{bmatrix}
\begin{bmatrix}
    \m x_l^q \\
    \m x_l^w \\
    \m x_s
\end{bmatrix}
=
\begin{bmatrix}
    \m O_{4 \times 1} \\
    \m C_{8 \times 1} \\
    \m O_{3 \times 1}
\end{bmatrix} \\
& [0.009,~ 0.1,~0,~0,~\m C^{\mr T},~0,~0,~0]^\mr{T} \leq \m x \leq \notag \\
& [0.027,~0.3,~30,~30,~\m C^{\mr T},~50,~0.45,~300]^\mr{T} \label{equ:constraint_vtilted},
\end{align}
\end{subequations}
where $\m C$ is a constant vector and each entry of the vector follows the aforementioned definition of $\m x$ for this problem. 

This is a problem with a relatively short decision space, in which equifinality effects are hypothesized to be minimized.

\subsection{Numerical Case Study 2 - Wooden-Board Catchment - Pollutant Concentration}
The objective of this numerical case study is to test the ability of the optimization model to predict the salt concentrations at the outlet of the catchment, the initial salt mass and the wash-off parameters of the model. This is a fairly more complex optimization problem if not only the water quantity parameters are required to calibrate, but also the water quality ones. In addition, the water quality parameters have a wider sensitivity as shown in \citep{gomes2023hydropol2d}. In this problem, we assume that the water quantity parameters (i.e., $n$ and $h_0$) are already calibrated \citep{zhang2020physically}, so that the decision vector for this problem is $\m x_l^w = [C_{1,1},~C_{2,1},~C_{3,1},~C_{4,1}]^{\mr T}$. To set only the water quality variables as the decision variables, we add an equality constraint in Eq.~\eqref{equ:equality_constraint}. In this problem, we only choose RMSE as the fitness function. 

This case study is a controlled experiment in a wooden-catchment as shown in Fig.~\ref{fig:case_study}b). The wooden-board has 4.5~$\mathrm{m^2}$ and 300 cells. The initial mass of the solute is 125 g and it is assumed that it is uniformly distributed in the catchment area \citep{zhang2020physically}. However, in this paper, we do not assume that the initial solute mass is known, and we let the model search for the near-optimal solute mass considering $C_1$ and $C_2$ as decision variables. Naturally, this controlled experiment is not a direct case of applying the build-up equation that calculates the available mass of the pollutant in terms of the ADD, as shown in Eq.~\eqref{equ:buildup}. However, fixing $\mr{ADD = 10~days}$, for example, we can estimate $C_1$ and $C_2$, calculate the initial build-up by solving Eq.~\eqref{equ:buildup} and compare with the initial mass of 125 g known from the experiment. Ultimately, what matters for HydroPol2D is the initial pollutant mass available in each cell of the domain. The reason we consider build-up as a function of ADD and LULC is that in most cases, the initial pollutant mass varying cell-by-cell is unknown and these direct measurements are either intractable, hardly ever available, and would result in an intractable decision-space if all cells are treated individually in the optimization problem. The parameter ranges were estimated from a 60\% variation from the previous calibrated ones, assuming that the initial pollutant mass was 125 g \citep{gomes2023hydropol2d}. In this Numerical Case Study, we want to answer the following question:

\begin{itemize}
    \item Q2: \textit{Assuming the water quantity parameters known, can the algorithm find the initial mass of salt (build-up model parameters) and the wash-off parameters to match with the observed pollutograph at the outlet?}
\end{itemize}

To answer this question, we build a formal optimization problem for this case study that can be written as follows.

\begin{subequations} \label{equ:opt_wooden}
\begin{align}
\min_{\m x} ~     O_f =  & ~\mr{RMSE}\\
&  \\
\textrm{s.t.}~~ &\mathrm{HydroPol2D~Dynamics}~\mr{in}~\mr{Eqs.}~\eqref{equ:explicit_scheme_overland}-\eqref{equ:pollutant_finite}
 \\
&\begin{bmatrix} 
    \m I_2 & \m O_{2 \times 7} \\
    \m O_{7 \times 2} & O_{7 \times 7}
\end{bmatrix}
\begin{bmatrix}
    n_1 \\
    h_{0,1} \\
    \m x_l^w \\
    \m x_s
\end{bmatrix}
=
\begin{bmatrix}
    0.04 \\
    0 \\
    \m O_{7 \times 1}
\end{bmatrix} \\
&\begin{aligned} &[0.04,~0,~166.71,~0.51,~5422.10,~0.14612,~0,~0,0]^\mr{T} \notag \\
& \leq \m x \leq \notag \\&[0.04,~0,~388.99,~1.19,~12651.57,~0.34096,~0,~0,0]^\mr{T},
\end{aligned}
\end{align}
\end{subequations}

\subsection{Numerical Case Study 3 - Gregorio Catchment in Sao Carlos / Brazil}

The Gregório catchment is located in the municipality of São Carlos in the state of São Paulo, Brazil. The climate in the state of São Paulo is influenced by Atlantic Tropical and Continental and Atlantic Polar air masses, complemented by Continental Equatorial air masses coming from the Western Amazon. The months with the largest rainfall events are in summer, from October to March, and the dry weather period varies from April to September in winter. The average annual precipitation of the city of São Carlos is approximately 1492 mm \citep{inmet2022} and the city has been prone to critical rainfall events yearly \citep{abreu2019quantificaccao}. The catchment area is 18.64~km\textsuperscript{2},  the length of the main channel is 8.6 km and its morphological characteristics indicate an elongated to strongly elongated catchment, which presents a compactness coefficient ($C_c$) of 2.030, a circularity ratio ($R_c$) of 0.120 and a form factor ($R_f$) of 0.289, as shown in Fig.~\ref{fig:Gregorio_Catchment}. Although the morphometric characteristics would indicate a resilient catchment to floods, the large impervious rate, mean slope, and the channelization of the main creek increase the vulnerability of the area in terms of floods. 

To perform hydrodynamic modeling, we built maps of Digital Elevation Model (DEM), Land Use Land Cover (LULC) and Soil Texture as presented in Fig.~\ref{fig:DEM_LULC_SOIL}. Due to the lack of high-resolution data in the catchment, we use freely available worldwide datasets for all input maps; therefore, the methods applied here are replicable in other poorly-gauged catchments \citep{gomes2023hydropol2d}. However, when available, higher resolution maps can be used.

The pedology of the catchment is composed of yellow-red latosoil (YRL) and small areas of purple latosoil (PL) \citep{ibge2022}. The soil texture within the catchment can be classified into medium and clayey texture \citep{ibge2022} (see Fig.~\ref{fig:DEM_LULC_SOIL}).  The headwaters of Gregorio catchment  remains relatively undeveloped, with a predominance of pervious areas with crops, grass, and shrub areas. Downstream the creek, impervious rates dramatically increase with the urbanization, which almost makes the catchment impervious towards the outlet, as presented in Fig.~\ref{fig:Gregorio_Catchment} and Fig.~\ref{fig:DEM_LULC_SOIL}. This catchment covers large proportion of the urbanized area in the municipality of São Carlos with the most commercial activities of the city being carried out in this area. For this reason, due to the climatic and hydraulic characteristics of the catchment, floods are constantly recorded, especially in the summer. A recent flood picture is shown in Fig.~\ref{fig:Gregorio_Catchment} in the Local Market \citep{abreu2019quantificaccao,sarmento2020using}.

The specific question of this Numerical Case Study is:
\begin{itemize}
    \item Q3: \textit{Given the reality of scarce data in poorly-gauged catchments, can the algorithm find the near-optimal hydrodynamic parameters, within physical limits, to match with the observed hydrograph? Can this set of parameters be used to estimate catchment-scale information?}
\end{itemize}

\subsubsection{DEM - Preprocessing}
A 30-m DEM might be considered high-resolution for rural catchments. For urban areas, however, the complexity of the built environment with detailed infrastructure would require a more detailed resolution. Information in such detail is typically unavailable in developing countries. Nonetheless, poorly-gauged areas are usually those that often suffer from floods \citep{fava2020improving}. Raster-based flood routing models are, therefore, affected by DEM quality. Typically, DEMs are required to be hydrologically corrected, ensuring that the flow directions are continuous and connected toward the outlet. However, especially in urban areas with bridges, culverts, and stormwater reservoirs, DEMs usually have to be burned to allow proper flow directions and connection. Herein, we provide 4 algorithms to treat low-quality DEMs (i.e., gaussian filter \citep{young1995recursive}, constrained regularized smoothing of the channel length profile (CRS) \citep{schwanghart2014topotoolbox}, and reduction of DEM elevation to consider water surface depths based on \citep{de2013large}. These methods are detailed in the Supplemental Material.
\subsubsection{Data Collection}
\begin{figure*}
    \centering
    \includegraphics[scale = 0.30]{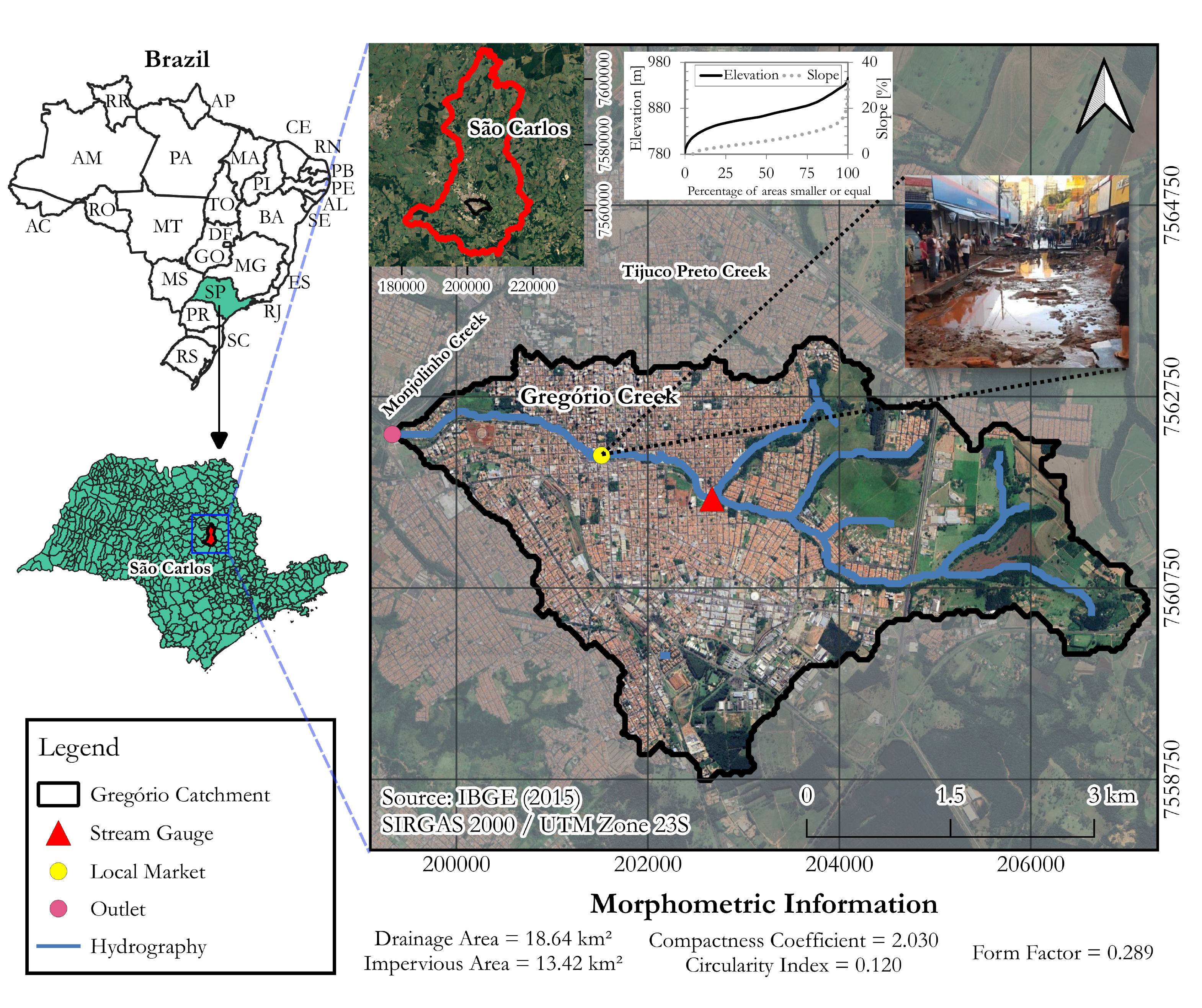}
    \caption{Gregorio Catchment location map with hypsometric curves of elevation and slope, and a figure of the flood-related impacts in the Local Market point of Sao Carlos. }
    \label{fig:Gregorio_Catchment}
\end{figure*}

Rainfall intensity and depth of the water surface are recorded in a limited way each minute and upscaled to 5-min intervals, and the rainfall and stream gauge station is shown in Fig.~\ref{fig:Gregorio_Catchment}. From the monitoring campaign provided in \citep{souza2008drenagem}, only one event had a sufficiently large rainfall volume and quality discharge observations. To collect data, a Campbell Scientific® CR10 station was installed and calibrated to record the data to be collected after each rainfall event \citep{souza2008drenagem}. A calibrated rating curve \citep{gomes2023rating} converts water depth into flow discharge by the following relationship \citep{lima2007analise}:
\begin{equation} \label{equ:rating_curve}
    Q(h) = 8.278 h^{2.2517},~r^2 = 0.99,
\end{equation}
where $Q$ is the observed flow discharge at the gauge station, and $h$ is the measured water depth taken from the channel invert.

\begin{figure*}
    \centering
    \includegraphics[scale = 0.32]{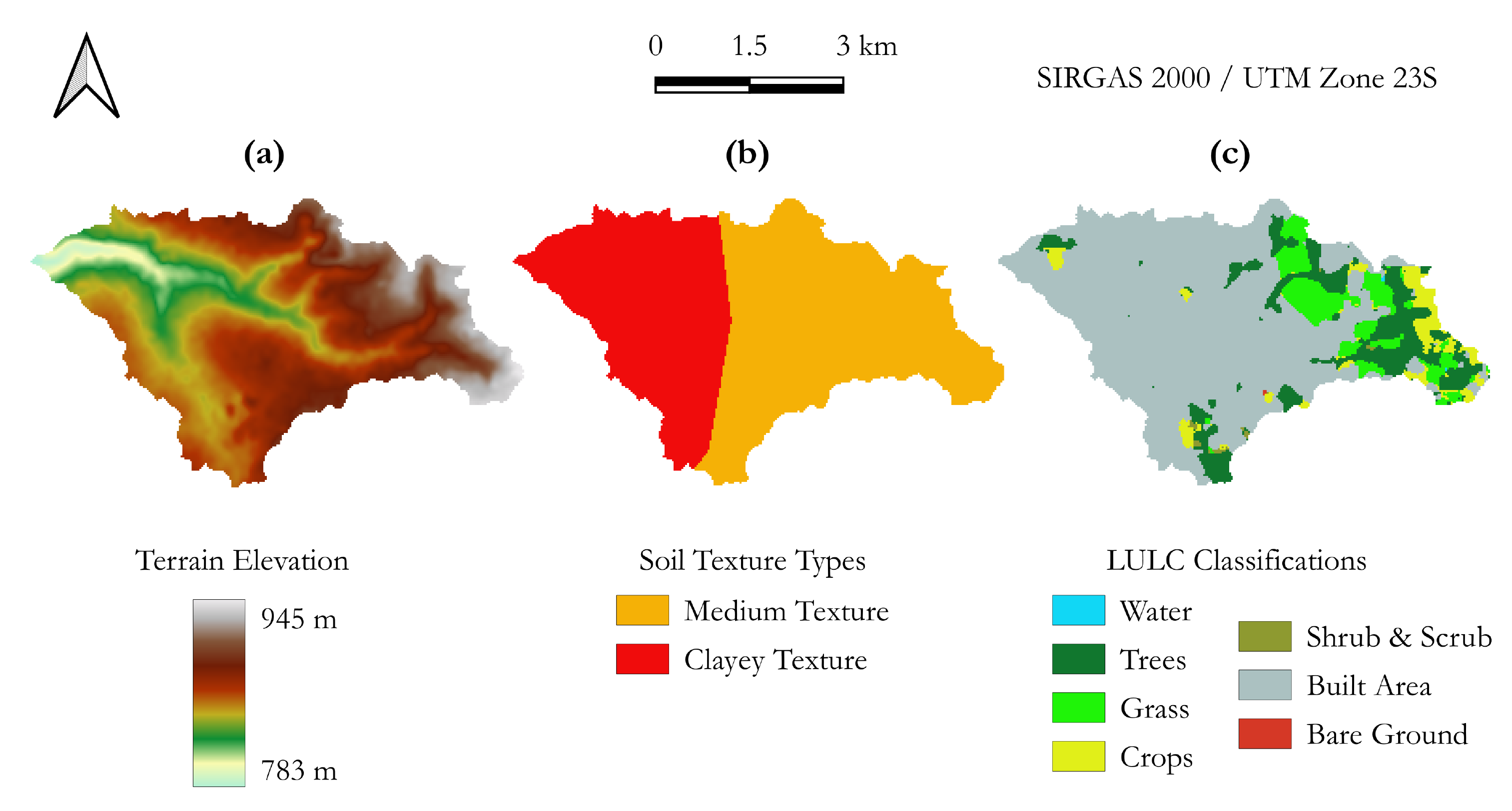}
    \caption{Input maps to the hydrodynamic simulation in HydroPol2D for the Gregorio catchment, where a) is the digital elevation model, b) is the soil texture map, and c) is the land use and land cover map.}
    \label{fig:DEM_LULC_SOIL}
\end{figure*}

The recorded level was converted into flow discharges using Eq.~\eqref{equ:rating_curve} and used for the calibration of HydroPol2D.

\subsubsection{Initial Conditions}
We assumed the initial soil moisture in the soil calculated with the cumulative rainfall prior to the event coupled with the SCS-CN \citep{scs1986urban} infiltration model. A Curve-Number map was developed by \citep{souza2008drenagem} and used to estimate spatial infiltration in pervious areas. For the initial water surface depth, previous modeling results indicate that no warm-up is necessary, and current visits to the study area indicate a minimum effect of non-hortonian flows. It is also seen from the observed hydrographs that the initial flow is null, indicating an intermittent creek.

\subsubsection{Boundary Conditions and Running Control Parameters}
The model is simulated with a space-invariant and time-variant rainfall hyetograph distributed to all cells of the grid. In addition, we assume a gradient outlet boundary condition at the outlet with the friction slope $s_f = 0.02~\mathrm{m/m}$. Although we assume normal flow at the catchment outlet, the flow is considered transient in the gauging station as it is an internal node of the domain (see Fig.~\ref{fig:Gregorio_Catchment}). To guarantee numerical stability, we define minimum and maximum time steps of 0.1 and 5~sec, respectively, and we set HydroPol2D model to change time-steps each 60 seconds of the simulation time. We use a Courant number bound of 0.4, such that time-steps are adapted to match this input \citep{gomes2023hydropol2d}. The model is run for 120 min of simulation and point and raster results are retrieved each 5-min.

\subsubsection{Parameter Ranges}
The parameter ranges used for calibration and for the construction of the calibration optimization problem of Eq.~\eqref{equ:optimization_problem} are given in Tab.~\ref{tab:LULC_Based_parameters} and Tab.~\ref{tab:SOIL_Based_parameters}. The physically bounds used in this paper were derived from the literature and recent published papers and manuals \citep{soliman2022assessment,rossman2010storm,brunner2016hec}.

\begin{table*} 
\centering
\begin{tblr}{
  cell{1}{1} = {r=2}{},
  hline{1,3,11} = {-}{},
}
Classification   & $n_{\mr{min}}$~      & $n_{\mr{max}}$~      & $n$          & $h_{\mr{0,min}}$ & $h_{\mr{0,max}}$~ & $h_{\mr{0}}$     \\
                 & $[\mathrm{sm^{-1/3}}]$ & $[\mathrm{sm^{-1/3}}]$ & $[\mathrm{sm^{-1/3}}]$ & ~[mm]  & [$\mr{mm}$]  & [$\mr{mm}$] \\
Water            & 0.0250      & 0.0400      & 0.0400      & 0.00      & 0.00       & 0.00  \\
Trees            & 0.0250      & 0.0400      & 0.0268      & 0.00      & 10.00      & 0.8259  \\
Grass            & 0.0200      & 0.0350      & 0.0244     & 0.00      & 5.00       & 0.2740  \\
Flooded Vegetation & 0.0250      & 0.0400     & 0.0381     & 0.00      & 10.00      & 1.2281  \\
Crops            & 0.0200      & 0.0350      & 0.0236      & 0.00      & 10.00      & 0.1564  \\
Scrub/Shrub     & 0.0300      & 0.0400      & 0.0358      & 0.00      & 10.00      & 7.6289  \\
Built Areas      & 0.0150      & 0.0300      & 0.0216      & 0.00      & 2.00       & 0.0625   \\
Bare Ground      & 0.0200      & 0.0300      & 0.0260      & 0.00      & 2.00       & 1.6101   
\end{tblr}
\caption{Parameter ranges and calibrated values for the LULC-Based parameters of HydroPol2D for Numerical Case Study 3}
\label{tab:LULC_Based_parameters}
\end{table*}

\begin{table*} 
\footnotesize 
\centering
\begin{tblr}{
  cell{1}{1} = {r=2}{},
  hline{1,3,5} = {-}{},
  hline{2} = {2-10}{},
}
Type & $k_{\mr{sat,min}}$ & $k_{\mr{sat,nax}}$ & $k_{\mr{sat}}$     & $\Delta \theta_{\mr{min}}$    & $\Delta \theta_{\mr{max}}$ & $\Delta \theta$    & $\psi_{\mr{min}}$~  & $\psi_{\mr{min}}$~  & $\psi$~  \\
               & [$\mr{mm.h^{-1}}$] &  [$\mr{mm.h^{-1}}$] & [$\mr{mm.h^{-1}}$] & [-] & [-] & [-] & [$\mr{mm}$] & [$\mr{mm}$] & [$\mr{mm}$] \\
Medium & 1.00        & 10.00       & 1.14       & 0.25          & 0.60    & 0.29         & 0.00      & 230.00    & 33.56      \\
Clayey & 0.20        & 10.00       & 6.23        & 0.25          & 0.60   & 0.40       & 0.00      & 312.50    & 281.10      
\end{tblr}
\caption{Parameter ranges and calibrated values for the SOIL-Based parameters of HydroPol2D for Numerical Case Study 3}
\label{tab:SOIL_Based_parameters}
\end{table*}

\subsubsection{Sensitivity Analysis}
A one-at-the-time sensitivity analysis is performed in the model to identify the most sensitive parameters prior to the automatic calibration procedure \citep{gomes2023modeling}. We define three output functions and calculate the variance of each perturbation in the decision variables in terms of the variance in the output functions. We evaluated the output variance of Peak Flow, Runoff Volume, and Time to Peak, as they are closely related to the hydrograph properties. We also evaluated flood areas. More details of the output functions are found in the Supplementary Material. The parameters ranged from 10\% to 190\% the baseline parameters, with 10\% intervals, and the baseline parameters are defined as the average of the parameter ranges presented in Tables \ref{tab:LULC_Based_parameters} and \ref{tab:SOIL_Based_parameters}.

\subsection{Numerical Case Study 4 - Exploring Equifinality}
In this case study, our objective is to explore the parameter equifinality problem in the calibration of the hydrological and hydrodynamic model. Equifinality tends to decrease with the number of observations and with the decrease in the model parameters \citep{her2019uncertainty}. To this end, we create a synthetic case study without uncertainty in rainfall, initial soil moisture, and observed discharge, mimicking a perfect gauging system. Therefore, the error in the parameter measurements is only due to the lack of accuracy in the optimization solver, number of generations, population size, genetic algorithm properties that might change the behavior of the exploration of the decision space, and due to the conceptual model of HydroPol2D. The spatial-variability of rainfall is a challenge that could be also explored, but is out of scope of this paper.
To explore the equifinality problem in a scenario of certain rainfall and perfect measurements in the gauges, we formulate the following question.

\begin{itemize}
    \item \textit{Q4-1: How does parameter equifinality affect the calibration of HydroPol2D for different parameter ranges, number of events, magnitude of the rainfall intensity, and location of the gauging stations?}
\end{itemize}

Therefore, we assess the near-optimal calibrated parameters in an inverse problem using the V-Tilted catchment as a surrogate case study, varying the number of gauges, the number of rainfall events, and its intensities. This catchment is used as a virtual laboratory to test the hypothesis raised in this case study \citep{fatichi2016overview}. We altered the original watershed to have 3 classifications of soils and LULC, following the left hillslope (1), the middle channel (2), and the right hillslope (3), each of them with different $n$, $h_0$, $k_{\mr{sat}}$, $\Delta \theta$, and $\psi$. The left and right hillslopes can be classified into hydrologic units with the same roughness and infiltration properties. Consequently, having a gauge station in each of the hillslopes would possibly reduce the uncertainty in the parameters.

The problem has ten unknown parameters (i.e., five parameters for each hillslope) and five known parameters (i.e., the main channel parameters are kept constant). We apply the model to calibrate three different storms of $10.8~\mathrm{mmh^{-1}}$, $21.6~\mathrm{mmh^{-1}}$, and $32.4~\mathrm{mmh^{-1}}$ with 90-min duration and later we calibrate only using the first event of $10.8~\mathrm{mmh^{-1}}$. Detailed results of the modeling of each of the three rainfall events are presented in the Supplemental Material. The number of gauges (1 - outlet, 2 - left hillsope, and 3 - right hillslope) is combined, resulting in 7 possible cases (1-2-3, 1-2, 1-3, 2-3, 1, 2, and 3). The left and right gauges are defined by the channel neighborhood cell located at the half middle of the V-tilted length (i.e., $y = 500~\mathrm{m}$) spanned 20 m from the channel i.e ($x = 780~\mathrm{m}$, and $x = 820~\mathrm{m}$, see Fig.~\ref{fig:case_study}a)).

We formulate the calibration problem with a wide parameter range (see Supplemental Material) mimicking no prior knowledge of the system except by the input data that discretize the domain into 3 areas (i.e., left hilslope, channel, right hillslope). We compare the results of this case with a condition with more knowledge of the system, that is, the parameter range is half of the previous one, hence reducing the decision space.

The calibration of hydrological models is inherently multi-objective \citep{shafii2009multi}. For example, minimizing RMSE might give good objective function values, correctly matching the peaks, but might fail during the recession time, thus altering the overall mass balance that is accounted for in soil moisture, for example \citep{lindstrom1997simple}. To this end, we use two metrics as our composite objective function, that is, the NSE and the relative volume error. The NSE varies from $\-\infty$ to 1 and the relative volume error should be minimized, such that we would want to maximize the NSE while minimizing the volume error. By introducing a penalizing factor as a function of the relative volume error in the NSE, we seek solutions that have a good NSE and reduced volume errors. Therefore, the objective function also varies from $-\infty$ to 1, where 1 indicates a perfect NSE and no volume error.

To transform this hypothesis into a minimizing optimization problem, we assume that each gauge has the same importance (i.e., $\gamma = 1/n_g$, with $n_g$ as the number of gauges), assuming the NSE with weight $\alpha_1 = 1$ and the volume error with weight $\alpha_2 = 0.5$ as well, and we assume that each event also has the same importance (i.e., $\beta_j = 1/n_e~\forall~j$, with $n_e$ being the number of events). Therefore, we can write the objective function as \eqref{equ:objective_function} \citep{lindstrom1997simple}.

\begin{equation}
\small
    O_f =  \frac{-1}{n_e}\sum_{j=1}^{n_{e}}  \Bigr[\frac{1}{n_g}\sum_{i = 1}^{n_g} \Bigl (\alpha_1 \overbrace{\mr{NSE}^{\mr{i,j}}}^{f_1}  - ~\alpha_2\overbrace{\abs{\frac{\sum_{k=1}^{T_j}(Q_{\mr m}^{\mr{i,j,k}} - Q_{\mr{obs}}^{\mr{i,j,k}} )}{\sum_{k=1}^{T_j}(Q_{\mr{obs}}^{\mr{i,j,k}} )}}}^{f_2} \Bigr) \Bigl],
\end{equation}
where $i$ is the gauge index, $j$ is the event index, and $k$ is a time-step index.

The previous objective function attempts to maximize NSE while trying to maintain important hydrological features such as volume conservation \citep{gomes2023modeling,lindstrom1997simple}. The negative sign in the first fraction is to transform the objective function suitable for the minimization of the optimization problem.

The calibration of hydrologic-hydrodynamic models is a necessary but not sufficient condition to the application of the model under different input ranges. The validation process is usually performed with different hydrologic conditions, typically represented by storms different from those used for the calibration. In order to gain confidence in the estimated parameters, we provide a validation test under different storm volumes, intensities, temporal distributions, and volumes. To attach these issues, we formulate the following question.

\begin{itemize}
    \item \textit{Q4-2: Using only the observed data at the outlet, is it possible to obtain a sufficiently accurate model that can be used not only for the calibration events but also for the validation under different storms intensities,  durations and temporal distributions?}
\end{itemize}

For the purpose of answering Q4-2, we calibrate the model with only the outlet gauge as the source information for the optimization calibration algorithm. To ensure different rainfall characteristics, we change the durations and volumes, as well as the rainfall temporal distribution. The rationale behind this is to have rainfall events with 50 or 150\% values from the calibration events, whenever possible to represent relatively different conditions from the calibration phase. Therefore, we alter the durations from the 90-min rainfall duration used from the calibration events, resulting in rainfall durations of either 45 or 135 minutes. However, reducing the rainfall volumes to 50\% of the smallest rainfall event used for calibration would generate events without runoff. Therefore, for this case, we fix the intensity as 10.8~$\mr{mmh^{-1}}$ but change the duration of the rainfall. To consider the effect of unsteady-state rainfall, we use the Huff 1st quartile hyetograph \citep{huff1967time} as a proxy rainfall distribution to represent the temporal dynamics of the rainfall. 

\begin{table}
\centering
\begin{tblr}{
  cell{1}{1} = {r=2}{},
  hline{1,3,6} = {-}{},
  hline{2} = {2-6}{},
}
Classification  & $n$     & $h_0$   & $k_{\mr{sat}}$ & $\Delta \theta$ & $\psi$  \\
                & [$\mr{sm^{-1/3}}$] & [$\mr{mm}$] & [$\mr{mmh^{-1}}$] & [$-$]   & [$\mr{mm}$] \\
Left Hillslope  & 0.06  & 1    & 8    & 0.6    & 20   \\
Channel         & 0.15  & 0    & 0    & 0.1    & 0    \\
Right Hillslope & 0.015 & 4    & 2    & 0.15   & 100  
\end{tblr}
\caption{Known parameters of the inverse problem of Numerical Case Study 4.}
\label{tab:known_parameters_4}
\end{table}

\section{Results and Discussion}

\subsection{Numerical Case Study 1} \label{sec:results_ncs1}
The modeling results of the V-Tilted catchment calibration problem are presented in Fig.~\ref{fig:num1_num2_results}. The near optimal value of the decision vector is found after 20 generations with 100 population are $n_1 = 0.0132~\mr{sm^{-1/3}}$, $n_2 = 0.1703~\mr{sm^{-1/3}}$, $h_{0,1} = 0.02~\mr{mm}$ $k_{\mr{sat}} = 0$, $\Delta \theta = 0.07~\mr{cm^3cm^{-3}}$, and $\psi = 15.21~\mr{mm}$. The hydrographs at the catchment outlet are presented in Fig.~\ref{fig:num1_num2_results}d), where the best and worst individual's hydrographs are plotted for each generation. Some of the worst individuals had no outflow due to large values of initial abstraction and/or $k_{\mr{sat}}$, likely due to the decision space exploration of GA. However, as the generation moves, the worst individuals predict better hydrographs with $\mr{NSE}$ closing to unity as shown in Fig.~\ref{fig:num1_num2_results}e), that shows the performance of the best and worst individuals throughout the generations. In nearly 1 generation, it is possible to note that the best individual already gets good results for hydrological models (i.e., $\mathrm{NSE} > 0.85$) \citep{nash1970river}.

The best individuals rapidly move to NSE closer to the unit, even though the parameters are not 100\% correct. Some of the parameter ranges used in this case study had nearly a 200\% variation from the minimum and maximum values (e.g., $n_1$ and $n_2$, as shown in Eq.~\eqref{equ:constraint_vtilted}) and the model still had minor errors compared to the expected values, as shown in Fig.~\ref{fig:num1_num2_results}f). The model also predicted that no infiltration would occur in this catchment since the near optimal $k_{\mr{sat}}$ is $0~\mr{mmh^{-1}}$. However, it predicted some initial abstraction of 0.02~mm, but this value is nearly negligible. Overall, the optimization problem resulted in a near-optimal solution that, at least for hydrological purposes, is a sufficient and physically based solution. It preserves the peak flow and overall shape of the hydrograph and has an optimal NSE compared to the outlet hydrograph. However, relying solely on NSE might produce great solutions, with feasible parameter estimations within the parameter boundaries but may be affected by equifinality, as shown in Fig.~\ref{fig:num1_num2_results}f), where it is noted that the Manning's roughness coefficient had a range of approximately 20\% around the expected values.

\subsection{Numerical Case Study 2}

\begin{figure*}
    \centering
    \includegraphics[scale = 0.45]{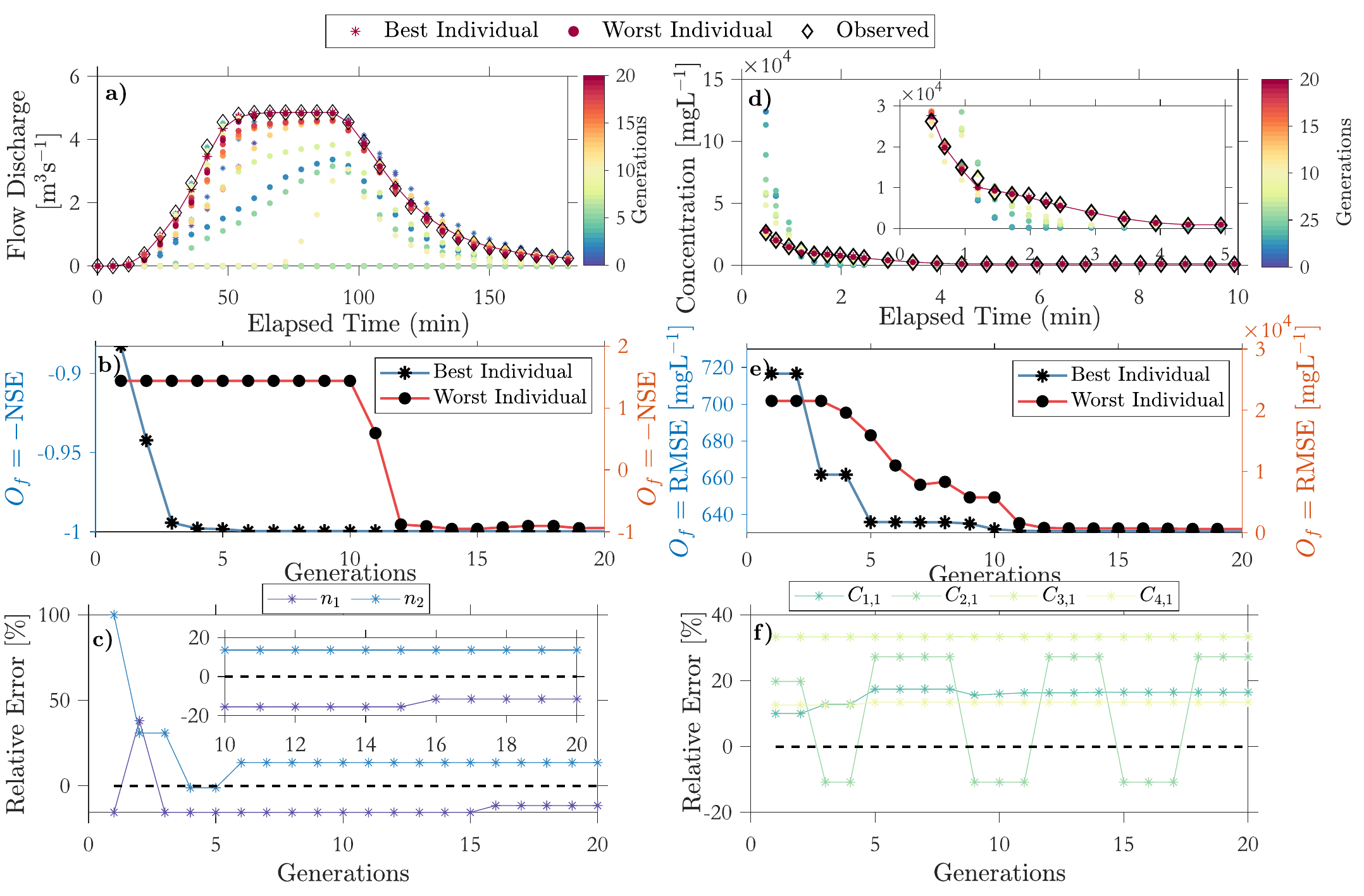}
    \caption{Modeling Results of Numerical Case Studies 1 and 2.  Parts a) - c) are the hydrograph, objective function chart, and relative error of parameters for the V-Tilted catchment. Parts d) - f) represent the pollutograph, objective function chart, and relative error chart for the Wodden-Plane catchment. Only the best and the worst individuals of each generation are plotted in a) and d). Only the parameters of the best individuals are plotted nin c) and f).}
    \label{fig:num1_num2_results}
\end{figure*}

The modeling results of the Wooden-Plane catchment are presented in Fig.~\ref{fig:num1_num2_results}(a)-(c). In this problem, the RMSE was chosen as the objective function and the nearly-optimal objective function value was approximately 630~mg/L. For other modeling pollutants, such as copper, zinc, or phosphate, a RMSE of this magnitude would result in an inaccurate model \citep{batalini2021evaluating}; however, we are modeling salt concentrations that had mean concentrations of approximately $30.000~\mr{mg/L}$, as shown in Fig.~\ref{fig:num1_num2_results}a). The goodness of fitness can also be visualized in the inserted chart in Fig.~\ref{fig:num1_num2_results}a), where the model nearly predicted the same concentrations as the observations.

The values of the objective function for the best and worst individuals of each generation are also shown in Fig.~\ref{fig:num1_num2_results}b). Results are already relatively good for the 1st generation and finds a near-optimal plateau after the 5th generation. Although the optimization model found good results for fitting the observed concentrations, it came at the cost of estimating a larger mass of salt in the beginning of the simulation. The overprediction of $C_1$ and $C_2$ can be seen in Fig.~\ref{fig:num1_num2_results}c), where $C_1 = 323.68$, $C_2 = 1.081$,~$C_3 = 12.045,03$,~$C_4 = 0.2763$. In particular, the combined values of $C_1,~C_2,$ and $\mr{ADD}$ would result in an initial salt mass of 145 g, which is approximately 16\% more than the value reported by \citep{hong2019physically}. The black dashed line in this figure is the result of the same problem, using the same model (HydroPol2D), calibrating for $C_3$ and $C_4$, but assuming the initial mass of 125 g \citep{gomes2023hydropol2d}. It is inferred that a larger mass was expected with a larger washing capacity, since all water quality parameters were larger for the simulations presented in this numerical case study.

\subsection{Numerical Case Study 3}
\begin{figure*}
    \centering    \includegraphics[scale = 0.65]{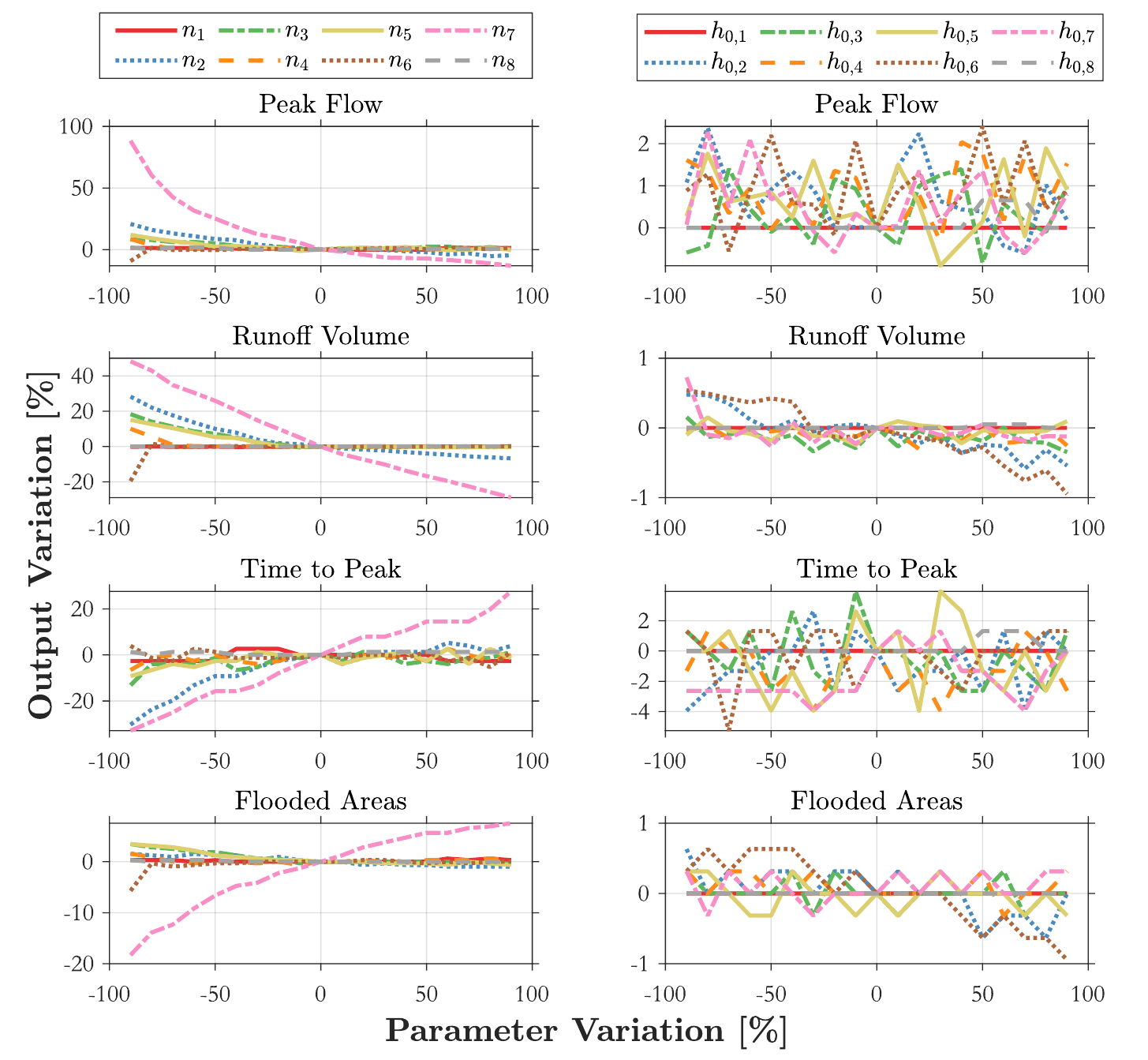}
    \caption{One-at-the-time sensitivity analysis of the LULC-Based parameters using the average parameters from the parameter range presented in Tab.~\ref{tab:LULC_Based_parameters} and Tab.~\ref{tab:SOIL_Based_parameters} for Numerical Case Study 3. LULC-Based subscripts 1 = water, 2 = trees, 3 = Grass, 4 = Flooded Vegetation, 5 = Crops, 6 = Schrub/Scrub, 7 = Built Areas, and 8 = Bare Ground.}
    \label{fig:LULC_Sensitivity}
\end{figure*}

\begin{figure*}
    \centering    \includegraphics[scale = 0.6]{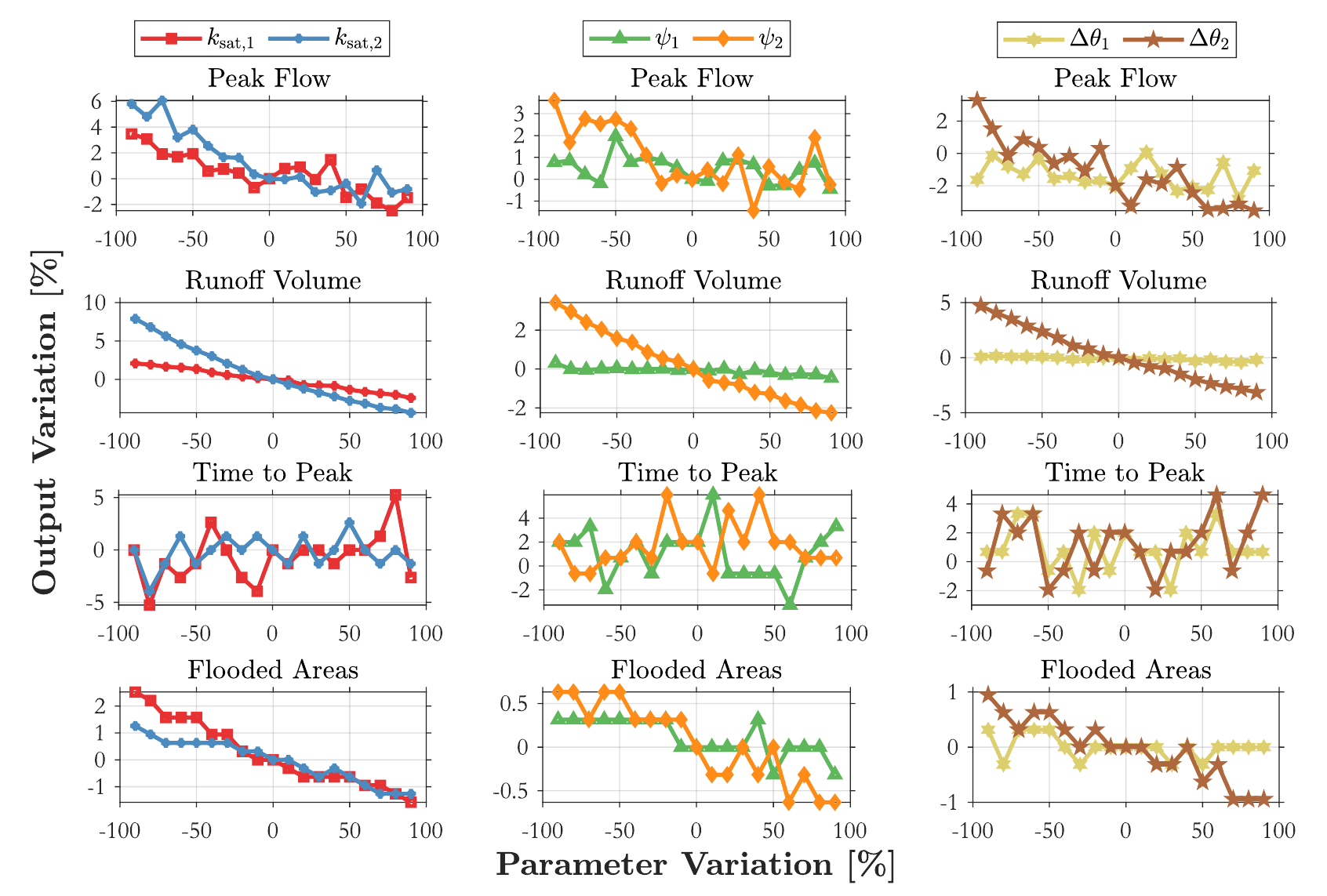}
    \caption{One-at-the-time sensitivity analysis of the SOIL-Based parameters using the average parameters from the parameter range presented in Tab.~\ref{tab:LULC_Based_parameters} and Tab.~\ref{tab:SOIL_Based_parameters} for Numerical Case Study 3. Soil-Based subscripts 1 = Medium and 2 = Clayey.}
    \label{fig:SOIL_Sensitivity}
\end{figure*}

The model has 8 LULC and 2 Soil classifications, resulting in 22 hydrologic-hydrodynamic parameters (16 from LULC and 6 from the soil parameters). Using the average of the parameter range presented in Tab.~\ref{tab:LULC_Based_parameters} and Tab.~\ref{tab:SOIL_Based_parameters}, the one-at-the-time sensitivities of these parameters are depicted in Figs.~\ref{fig:LULC_Sensitivity} and \ref{fig:SOIL_Sensitivity}. The average of the parameters might be the baseline used when only the parameter ranges are known. The output functions used for this evaluation are mainly related to the hydrograph shape. We used peak flow, runoff volume at the end of the event, and time-to-peak variances as hydrograph shape evaluation functions. Furthermore, we evaluate the sensitivity of the parameters to flood areas (i.e., areas with maximum flood depth greater than 0.5 m). From Fig.~\ref{fig:LULC_Sensitivity} it is observed that the most sensitive parameters in terms of hydrograph shape are the Manning's roughness coefficient of the Built Areas (i.e., the watershed has nearly 70\% of built areas), followed by those in the areas of trees. A reduction in roughness is fairly more expressive than an increase in terms of peak flow. However, for runoff volume, time to peak, and flooded areas, the Manning's variation seems to follow a linear relationship with these outputs.

It is interesting to note that reducing the $n$ values reduces the total volume that leaves the catchment at the end of the event and also decreases flooded areas, which does not necessarily mean that areas with risks of human instability would also decrease \citep{rotava2013simulaccao}. The larger velocities that resulted from the reduced Manning's coefficient might increase areas of instability risks. The initial abstraction ($h_0$) also had some sensitivity, but presented a very non-linear behavior for all output functions used, indicating that it could be assumed in some cases rather than calibrated. This non-linear behavior might be due to Eq.~\ref{equ:manning} from Algorithm \ref{alg:automata}, which allows storing water in cells if the surface runoff depths are smaller than or equal to $h_0$, while allowing infiltration. We hypothesize that $h_0$ would have more influence for values larger than 10~mm and this parameter can be used to represent the storage of low impact development (LID) facilities that do not change surface roughness, such as rain barrels. Parameters $n$ and $h_0$ can be proxy representations of LID facilities such as rain barrels (i.e., increasing $h_0$ in pixels), green roofs, permeable pavements, or bioretention systems, as they represent the storage and delay of the flood wave passing through cells \citep{damodaram2010simulation}.

These parameters associated with the infiltration parameters shown in Fig.~\ref{fig:SOIL_Sensitivity} can be used to assess the effects of retrofitting the catchment into a more sustainable scenario with green infrastructure \citep{mcclymont2020towards} or can also represent a scenario of increase in urbanization and hence represent the effects of post-development conditions \citep{gomes2023modeling}. Due to the limited area for infiltration in the catchment, the results presented in Fig.~\ref{fig:SOIL_Sensitivity} indicate that the soil properties have less influence than the roughness coefficients but greater influence than the initial abstraction, especially $k_{\mathrm{sat,2}}$, which can be seen from Fig.~\ref{fig:DEM_LULC_SOIL} that most of the pervious areas are derived from this type of soil.


\begin{figure*}
    \centering
    \includegraphics[scale = 0.53]{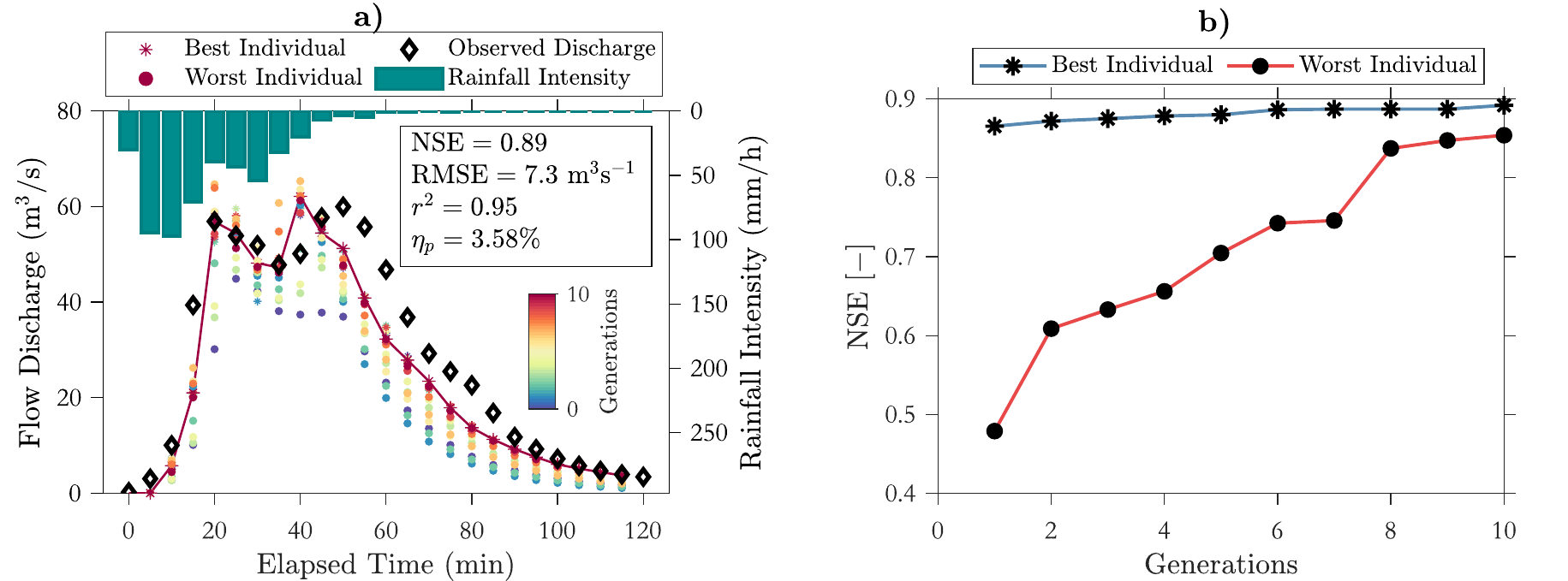}
    \caption{Calibration problem results with the catchment simulated with 30 m spatial resolution under an observed rainfall event. Part a) shows the hydrographs of the best and worst individuals, for each generation, as well as the rainfall intensity. Part b) shows the objective function (i.e., $\mathrm{- NSE}$) values for the best and worst individuals}
    \label{fig:hydrograph_gregorio}
\end{figure*}

The model calibration results are presented in Fig.~\ref{fig:hydrograph_gregorio}. Part a) shows the hydrograph for the best and worst individuals of each generation, while part b) shows the evolution of the objective function (i.e., - NSE) through the generations. The calibration results present a NSE = 0.89, $\mathrm{RMSE~=~7.3 ~m^3/s}$, and $r^2 = 0.95$, and $\eta_p = 3.58\%$. Although we ran the optimization algorithm for 10 generations, 100 population, it was fundamentally impossible to capture the second peak of the two-peak observed hydrograph, as well as the observed runoff volume in the falling limb of the hydrograph. Other objective functions could also be tested; especially that account for peak flows and runoff volumes, but for the sake of simplicity, we only used the NSE. Several factors might have influenced this behavior, and we hypothesize that the most important are the quality of the digital elevation model, the spatial variability of rainfall in the catchment, and the uncertainty in the rainfall measurements and transformation of stage into discharge, as well as the inherent uncertainty of the conceptual model of HydroPol2D. The calibrated parameters are shown in Tables \ref{tab:LULC_Based_parameters} and \ref{tab:SOIL_Based_parameters}.

\begin{figure*}
    \centering
    \includegraphics[scale = 0.85]{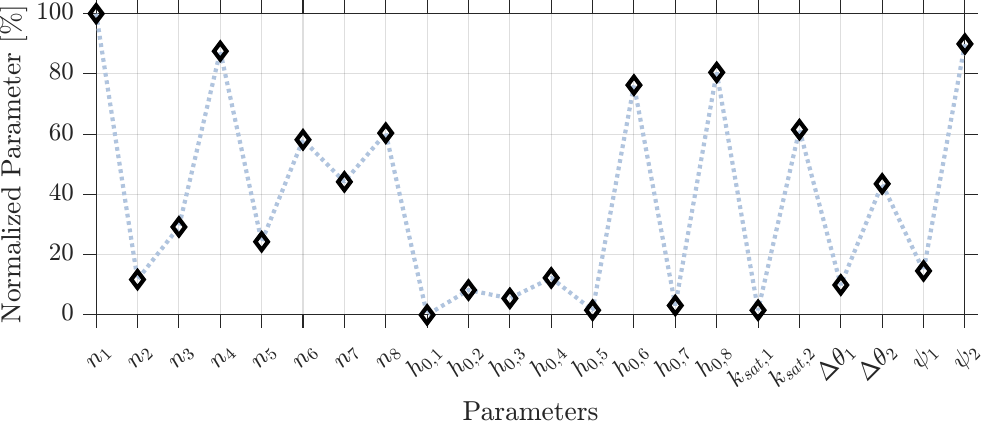}
    \caption{Near-Optimal normalized parameter, where 0 and 100\% are the boundaries of the decision vector $\m x$. LULC-Based subscripts 1 = water, 2 = trees, 3 = Grass, 4 = Flooded Vegetation, 5 = Crops, 6 = Schrub/Scrub, 7 = Built Areas, and 8 = Bare Ground. Soil-Based subscripts 1 = Medium and 2 = Clayey.}
    \label{fig:normalized_parameters}
\end{figure*}

As in Numerical Case Studies 1 and 2, after the first generation, the model performs similar to the last generation (i.e., NSE = 0.85), indicating that relatively fewer simulations can be required to reach accepted modeling results \citep{moriasi2015hydrologic}. As shown in Fig.~\ref{fig:normalized_parameters}, none of the parameters reached the boundaries of the range defined for the upper and lower bounds. However, a relatively large Manning's roughness coefficient is noted. Using a coarser DEM and filtering the DEM with Gaussian filters, CRS, and carving water surface depths in the channel, we hypothesized that the flow paths were shortened and smoothed in such a way that increasing $n$ was necessary. However, not including these filters might cause water ponding in areas with DEM noise. Furthermore, by using a 30-m DEM, the terrain details that would be captured with a relatively high resolution DEM and would eventually create larger flow paths (i.e., more contact area between the flow and the surface resulting in larger head losses) would have played a role in the relatively larger $n$ values. By having a shorter flow path (i.e., coarse resolution versus high resolution DEMs), for the same head loss, the $n$ values would have to increase.

\begin{figure*}
    \centering
    \includegraphics[scale = 0.70]{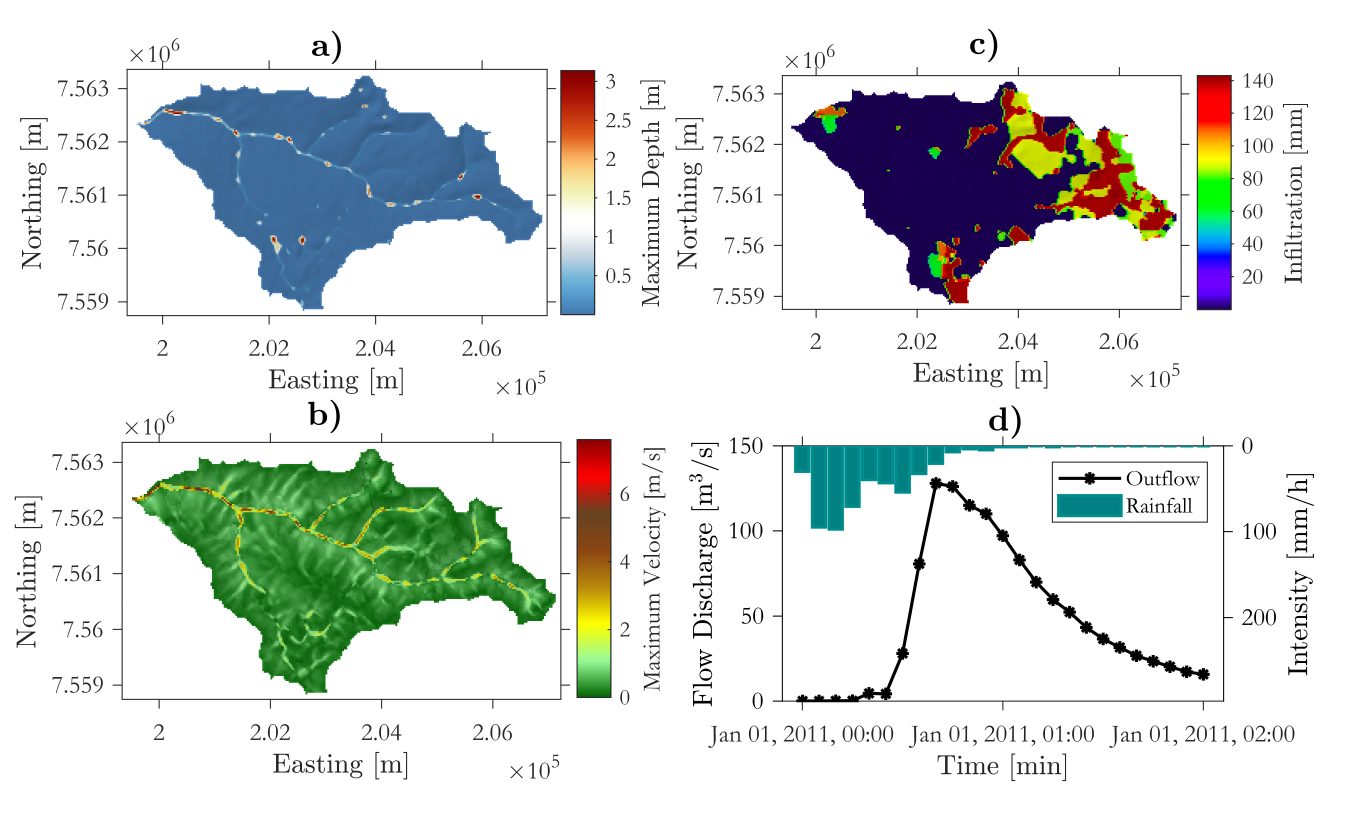}
    \caption{Some of the states modeled with HydroPol2D derived from a calibrated model with source point data used. Part a) is the maximum water surface depth, b) is the maximum overland flow velocity, c) is the infiltrated depth at the end of the event, and d) is the outlet hydrograph. The maps are projected in SIRGAS 2000 UTM 23 S.}
    \label{fig:states_hydropol}
\end{figure*}

Using calibrated parameters, HydroPol2D can be applied and determine flood maps, human instability maps, infiltration, and other components of the water balance as shown in \citep{gomes2023hydropol2d}. Fig.~\ref{fig:states_hydropol} shows the spatialized results derived from the calibrated model using the source point data from the observed gauge. It is possible to assess flood, infiltration, and velocity maps, as well as the outlet hydrograph. The Supplemental Material contains HydroPol2D outputs that can improve flood risk management if a calibrated model is available. 

\subsection{Numerical Case Study 4}

\begin{figure*}
    \centering
    \includegraphics[scale = 0.80]{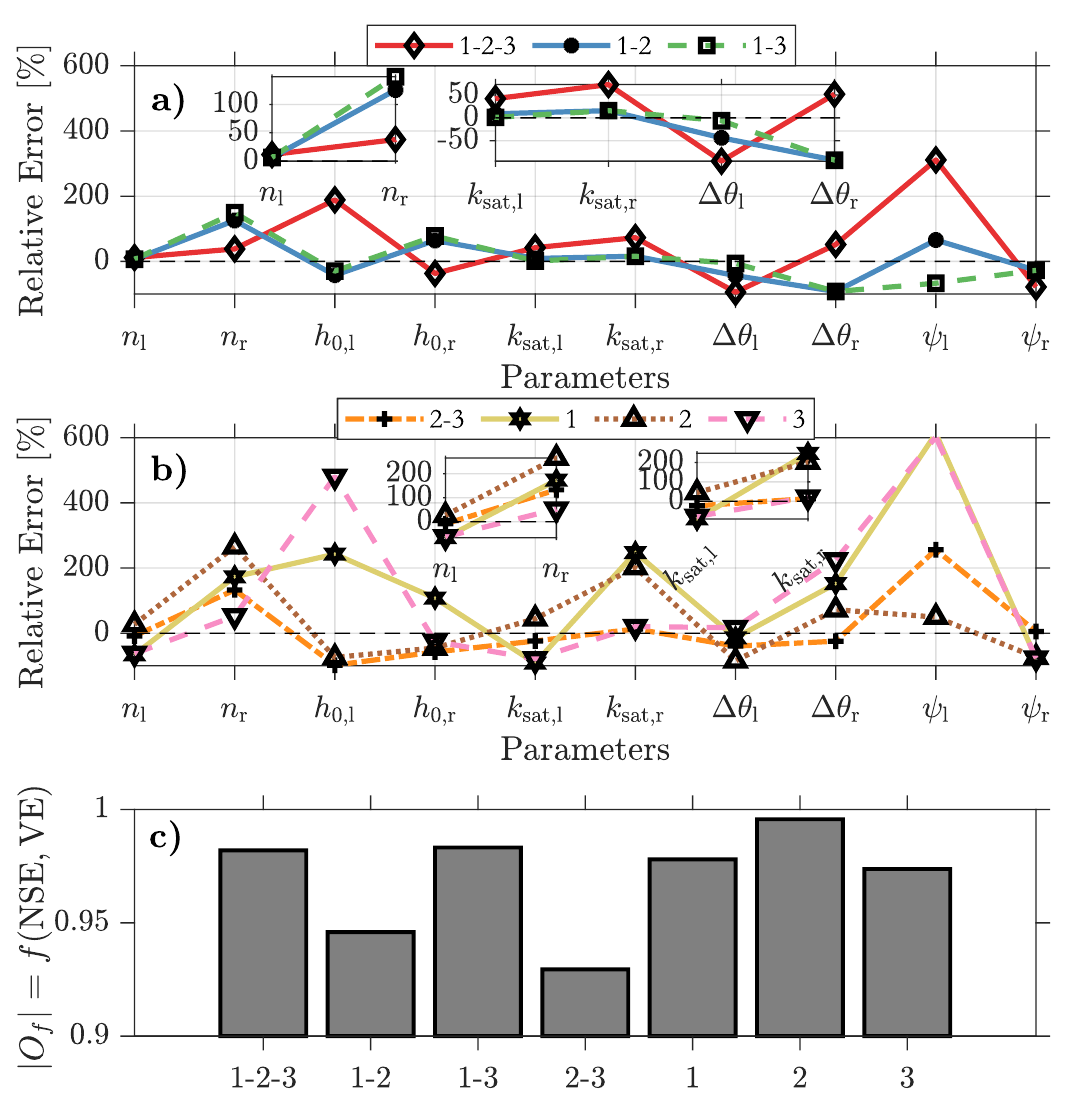}
    \caption{Relative Parameter Error for Numerical Case Study 4, assuming no prior knowledge of the parameter boundaries where 1 = Outlet, 2 = left gauge, 3 = right gauge. Black dashed lines are the expected values. Part a) is the relative error for cases where the outlet and at least one more gauge is observed and case, Part b) are single gauges or a combination of gauges that are not at the outlet and Part c) are the objective function values given by Eq.~\eqref{equ:optimization_vtilted}. All cases were simulated with 10 generations and 100 populations.}
    \label{fig:equifinality_no_PK}. 
\end{figure*}
The relative parameter error assuming a considerably wide parameter range that would represent no prior knowledge about the parameters is shown in Fig.~\ref{fig:equifinality_no_PK}. Overall, the algorithm can find suitable sets of $n$ and $k_{\mathrm{sat}}$, i.e., the most sensitive parameters using the outlet gauge in combination with some of the other gauges, but have larger errors in the other less sensitive Green-Ampt parameters and in the initial abstraction, as shown in Fig.~\ref{fig:equifinality_no_PK}a). These other parameters are less sensitive and could have been assumed otherwise without prejudice to the overall representation of the hydrology and hydrodynamics. 

When using gauges 2 or 3, that is, gauges from the left and right hillslopes, the error in the parameters of the opposite hillslope is relatively larger since the parameters were randomly chosen because no information about the discharges of that area is available, as shown in Fig.~\ref{fig:equifinality_no_PK}b). Using gauges 2 and 3 together, the error in the parameters is reduced since information from both hillslopes (i.e., areas with similar hydrologic characteristics) is available. Using only gauge 1 as the source of information for calibration, a good objective function value is found (see Fig.~\ref{fig:equifinality_no_PK}c)) but a very poor description of the parameters is found, as shown in Fig.~\ref{fig:equifinality_no_PK}b), indicating a high chance of equifinality if no minimum knowledge of the parameters is known. This result shows the importance of experts defining proper parameter ranges and that a previous, at least minimum, parameter range of the system is very important. All objective function performances could be considered feasible solutions with high fitness values if no prior knowledge of the system is known, as shown in Fig.~\ref{fig:equifinality_no_PK}c), increasing the equifinality issues.

\begin{figure*}
    \centering
    \includegraphics[scale = 0.85]{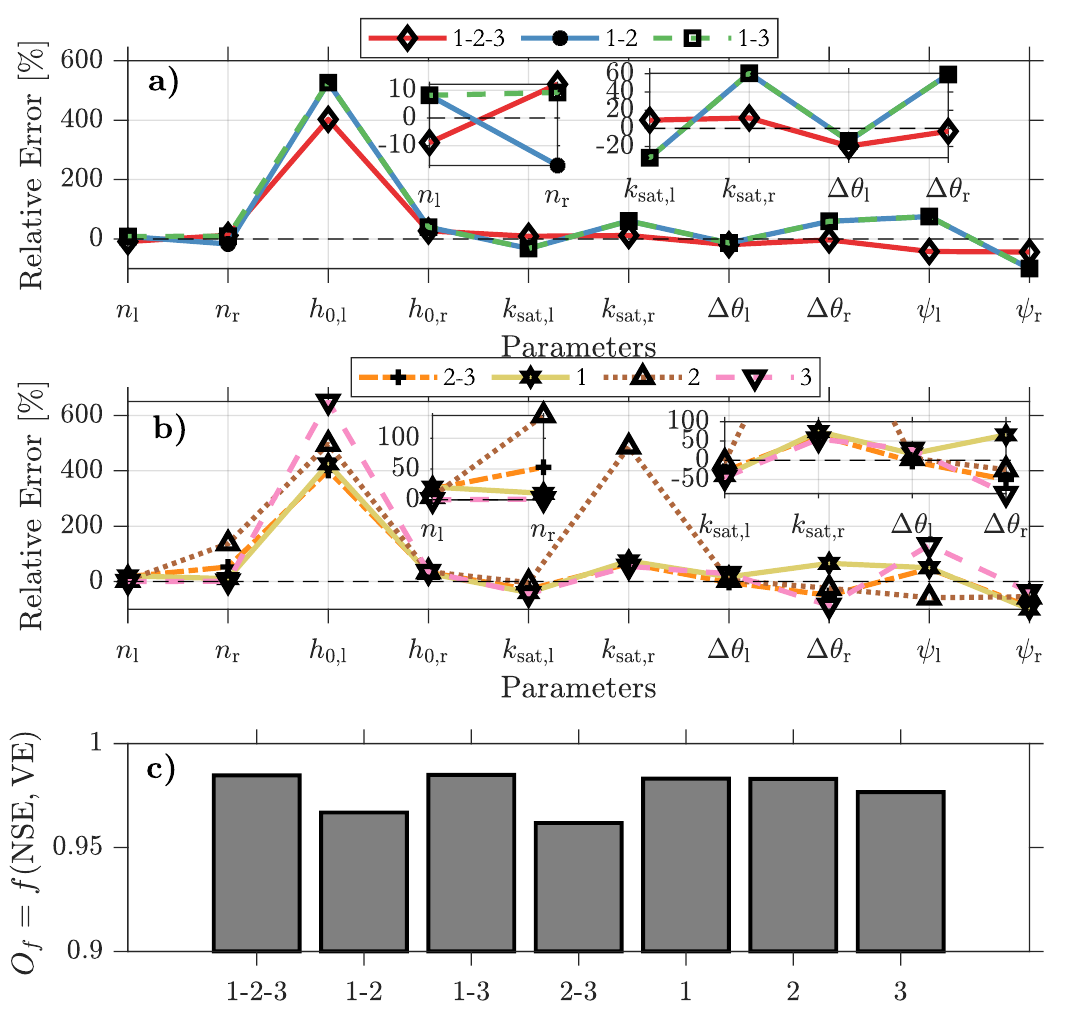}
    \caption{Relative Parameter Error for Numerical Case Study 4, assuming a prior knowledge, that is, half of the parameter range from Fig.~\ref{fig:equifinality_no_PK} of the parameter boundaries where 1 = Outlet, 2 = left gauge, 3 = right gauge. Black dashed lines are the expected values. Part a) is the relative error for cases where the outlet and at least one more gauge is observed and case, Part b) are single gauges or combination of gauges that are not at the outlet and Part c) are the objective function values given by Eq.~\eqref{equ:optimization_vtilted}. All cases were simulated with 10 generations and 100 population.}    \label{fig:prior_knowlegde_3events}
\end{figure*}

When comparing the performance of solutions without prior knowledge of the parameters with solutions with a smaller decision space, the equifinality tends to decrease, as shown in Fig.~\ref{fig:equifinality_no_PK} compared to Fig.~\ref{fig:prior_knowlegde_3events} and the values of the objective function are generally higher. This result indicates that a reduction in decision space to a more reduced space can substantially decrease equifinality. Some parameters less influential as $h_{0,l}$ had larger errors but little sensitivity and could have been assumed rather than calibrated.

The number of events in the calibration also plays an important role in the reduction of equifinality. By using a relatively small rainfall event that is not a runoff producing event in the left hillslope, no quality information is available for the calibration of the hydrodynamic and infiltration parameters of this hillsope. 
The left gauge in event 1, with 10.8~$\mathrm{mmh^{-1}}$ did not record runoff. Therefore, any combination of parameters such that all water infiltrates in the soil are a solution with full performance of the objective function. An infinite number of combinations of parameters would satisfy this condition (e.g., $h_{0,\mathrm{l}} > 16.2~\mathrm{mm}$, $k_{\mathrm{sat,l}} > 10.8~\mathrm{mmh^{-1}}$), leading to high equifinality due to poor gauging location and selection of the event for calibration. However, for the right gauge, runoff is observed, and the parameters can be relatively well estimated, although not perfectly. The problem of distributed physically-based modeling in small catchments with events that produce little or no runoff is a complex problem and models typically have lower performance for hortonian small flows \citep{senarath2000calibration,downer2004gssha}.

This idea is illustrated in Tab.~\ref{tab:only_one_storm}. Overall, by choosing only one poor event, the calibration performance is nearly optimal for all combinations of gauges, but the parameter estimation is faulty. Therefore, calibrating the model for longer hydrological periods or choosing a combination of events that would encompass relatively high and low flows is desirable to increase the available information and reduce parameter equifinality. However, the uncertainty in the boundary conditions and in the initial simulation values, especially the initial soil moisture \citep{senarath2000calibration}, is challenging. Even in a perfect virtual experiment without uncertainty in rainfall values, initial soil moisture, model boundary conditions, and perfect gauging data, the uncertainty in the parameters is substantially affected by a poor parameter range. 

One of the advantages of this calibration approach is the use of the model to calibrate the parameters using only the outlet data as the sole gauge, which would be the case of many poorly-gauged and flood-prone catchments such as the Gregorio Catchment. To this end, we use a relatively high optimization resource, that is, we run the optimization model for 40 generations and 100 population size and optimize Eq.~\eqref{equ:optimization_vtilted} using only the outlet as the observed gauge for all events available. The rationale behind using a larger number of generations is to extract the maximum resource of the single-point observed information, since it is only at one gauge. Using a larger number of generations would likely decrease the possibility of finding local optima in the optimization model. Therefore, the uncertainty would probably be due to equifinality since there is no uncertainty in rainfall and observations. 

\begin{figure*}
    \centering
    \includegraphics[scale = 0.62]{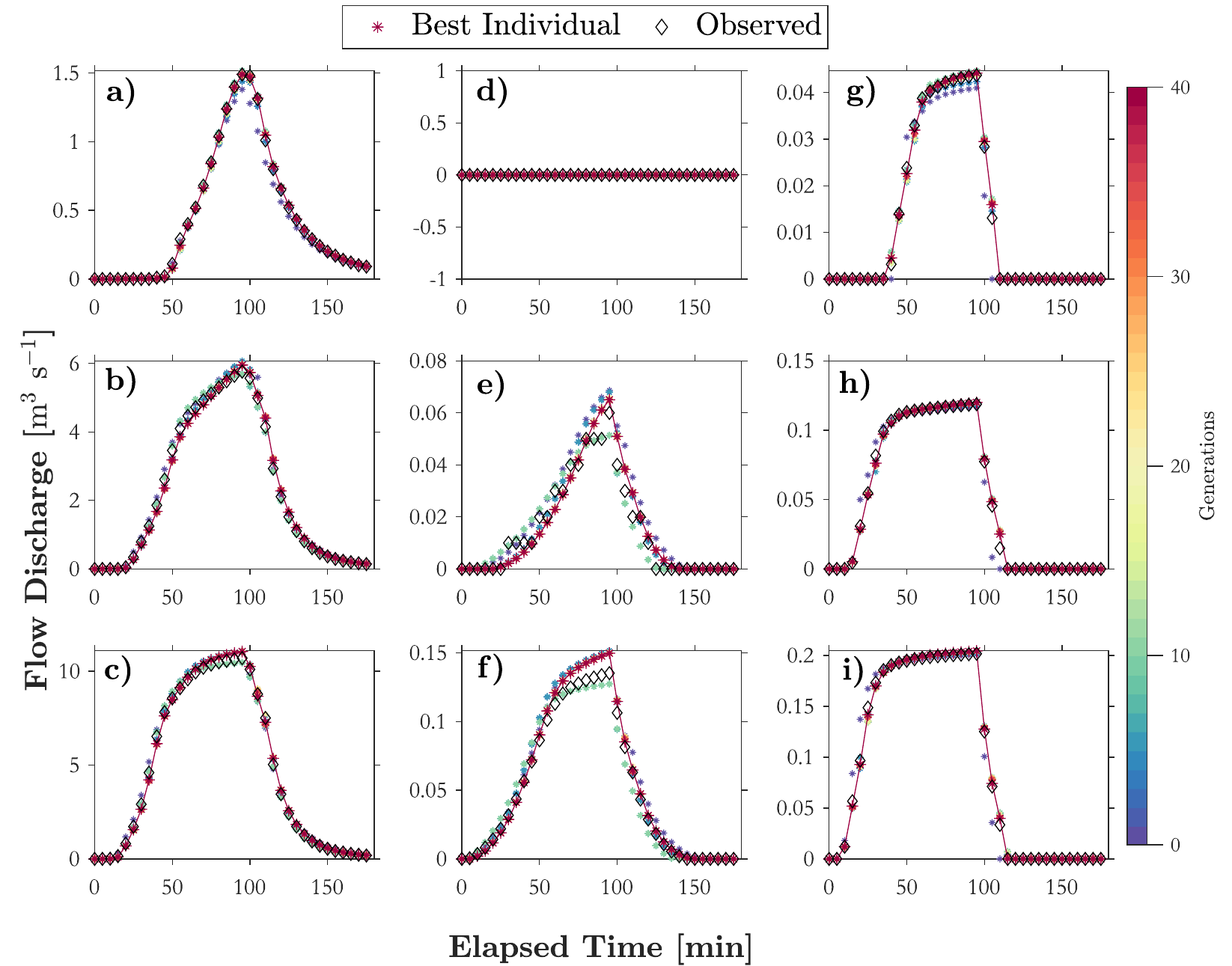}
    \caption{Automatic calibration results of considering the 3 events but only the outlet as the observable gauge. Therefore, the charts (d)-(i) are shown, but the modeled results were not considered in the calibration of the model and were simulated with the parameters that were obtained by calibrating the model only with the outlet gauge. The first row are the events of 10.8~$\mr{mm~h^{-1}}$, followed by 21.6~$\mr{mm~h^{-1}}$ and 32.4~$\mr{mm~h^{-1}}$. Parts (a)-(c) are results for the outlet, whereas (d)-(f) are from the left gauge and (g)-(i) are from the right gauge.}
    \label{fig:hydrograph_all_events}
\end{figure*}

In this analysis, we assume the initial abstractions of the left and right gauges as the correct parameters, since they do not play an important role in the hydrological response of the catchment, as mentioned above in this section. The other parameter ranges are the same as those used in the simulated cases with prior knowledge of the system (see Tab.~S1). The results in Fig.~\ref{fig:hydrograph_all_events} show the hydrographs for the outlet (a)-(c) and for the other gauges not considered in the calibration (i.e., left gauge (d)-(f), and right gauge (g)-(i)). Even calibrating with only the outlet of the catchment, the model can still find a reasonable, physically-based, and bounded parameter set, although the parameters are not equal as the one of the inverse problem. This result points to the scenario that, given a sufficient number of runoff-producing events and reasonable computational resources, it is possible to calibrate HydroPol2D only with data at the outlet and later use the calibrated model to derive important catchment response information such as those presented in Fig.~\ref{fig:states_hydropol}. The calibrated parameters of this analysis are $n_l = 0.0536~\mr{sm^{-1/3}}$, $n_r = 0.0168$, $k_{\mr{sat,l}} = 3.56
~\mr{mmh^{-1}}$, $k_{\mr{sat,r}} = 1.77~\mr{mmh^{-1}}$, $\Delta \theta_l = 0.625$, $\Delta \theta_r = 0.346$, $\psi_l = 91.36~\mr{mm}$, $\psi_r = 49.92~\mr{mm}$.

\begin{figure*}
    \centering
    \includegraphics[scale = 0.18]{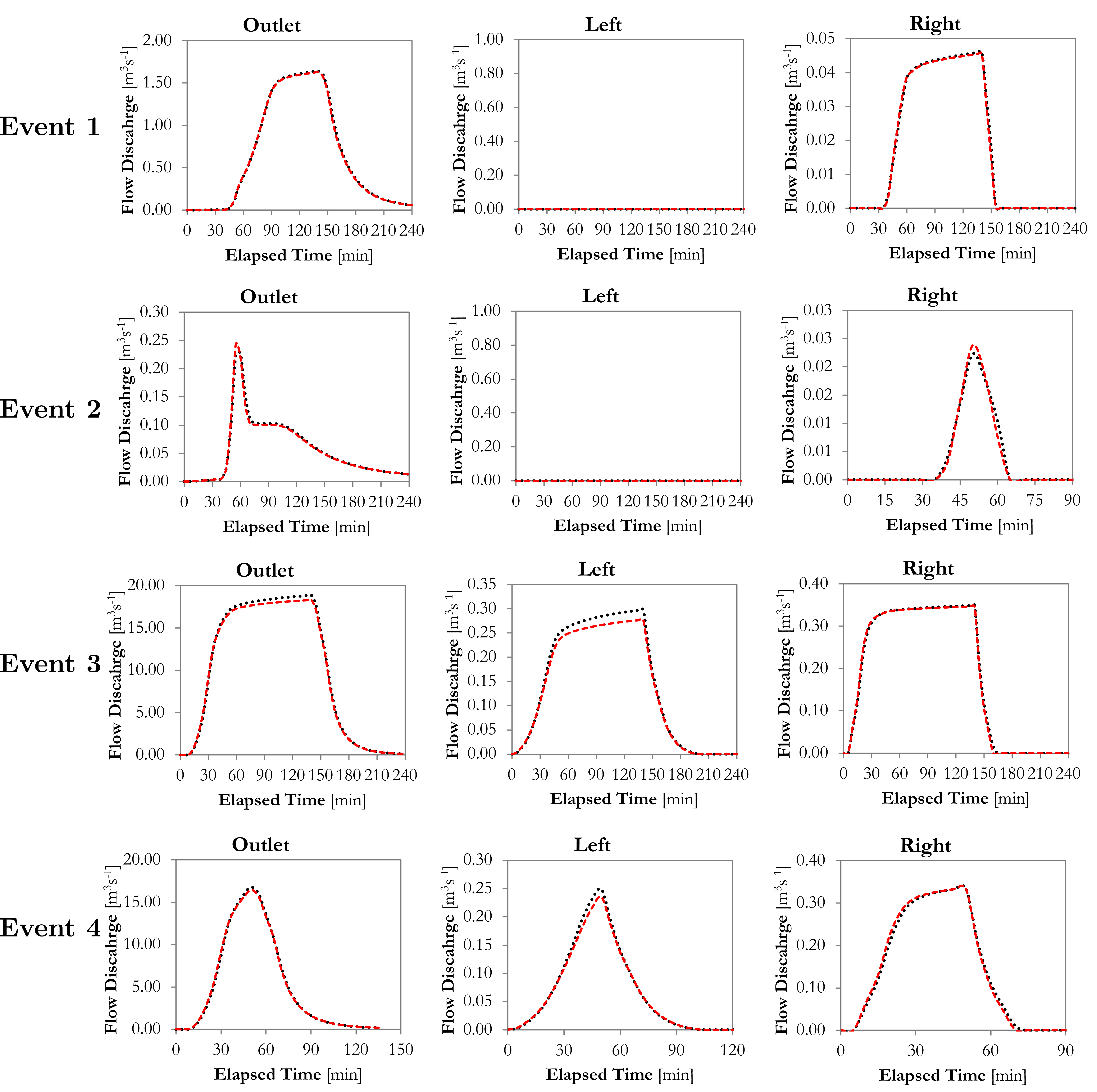}
    \caption{Steady-state rainfall validation hydrographs of Numerical Case Study 4,  for Events 1 to 4 described in Tab.~\ref{tab:validation}. Black dotted lines are modeled results with the calibrated model using only the outlet gauge data, and red dashed lines are the results with the parameters of the inverse problem.}
    \label{fig:event1_4_validation}
\end{figure*}

\begin{figure*}
    \centering
    \includegraphics[scale = 0.18]{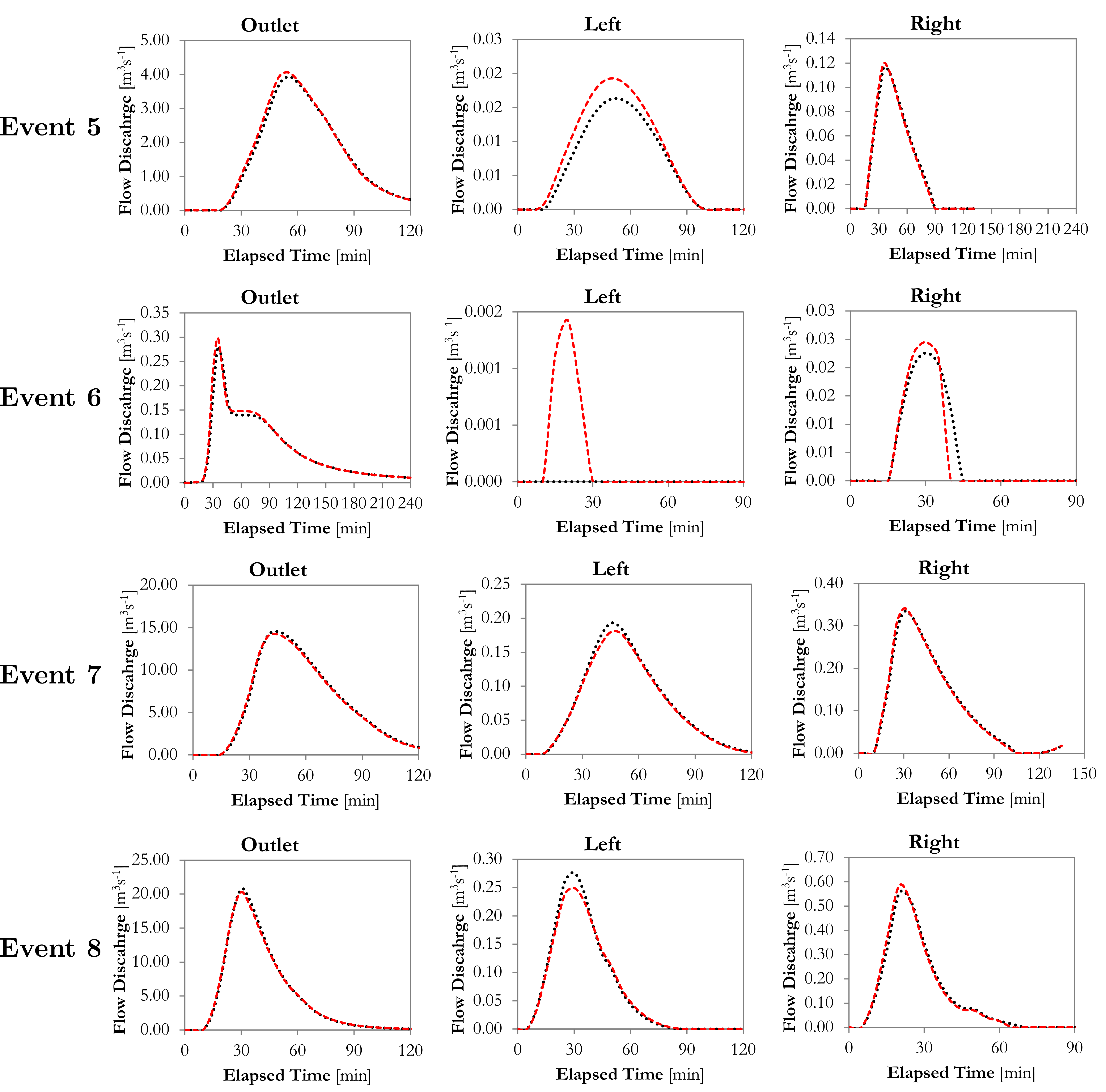}
    \caption{Unsteady-state rainfall results of Numerical Case Study 4,  for Events 5 to 8 described in Tab.~\ref{tab:validation}. The rainfall is simulated with Huff 1st quartile hyetograph. Black dotted lines are modeled results with the calibrated model using only the outlet gauge data, and red dashed lines are the results with the parameters of the inverse problem.}
    \label{fig:event5_8_validation}
\end{figure*}

By comparing the calibrated parameters with the ones of the inverse problem shown in Tab.~\ref{tab:known_parameters_4}, it is noticed that a trade-off between $k_{\mr{sat,l}}$ and $\psi_\mr{l}$ is found for the left hilslope. While $k_{\mr{sat,l}}$ decreases, $\Delta \theta_{\mr{l}}$ and $\psi_{\mr{l}}$ increase, counterbalancing the reduction in $k_{\mr{sat,l}}$. However, even though the parameter equifinality is evident, the model performance and the errors are visually minimal, as shown in Fig.~\ref{fig:event1_4_validation} and Figs.~\ref{fig:event5_8_validation}. In addition, the performance metrics are also accepted in most gauges, as shown in Tab.~\ref{tab:validation}.

For steady-state events, the model presented acceptable results for all events, with all volume errors smaller than 6\% and all NSE larger than 0.996. For unsteady-state hyetographs, as expected, the model presented a relatively reduced performance for the left gauge, especially for event 6, that is, the event with the smallest duration and volume. As shown in Fig.~\ref{fig:event5_8_validation}, event 6 generated a very small runoff rate that was observed in the inverse problem and not predicted by the calibrated model. In addition, for the right gauge, a relatively large volume error can be observed. For the outlet, however, the results are still quite accurate; although relatively faulty for the left and right gauges. Disregarding this event, the simulation results had volume errors smaller than 20\% and NSE larger than 0.992. Overall, using the calibrated parameters that were obtained only with the outlet gauge is sufficient to explain the events used for calibration and can accurately represent the hydrological response of events that are outside the hydrological characteristics of the events used for calibration. Even though some errors are found in the internal gauges, the performance of the model measured in the outlet can be considered very good for all validation events. 

\begin{table*}
\small
\centering
\begin{tblr}{
  cell{1}{1} = {r=2}{},
  hline{1,3,10} = {-}{},
}
Gauges & $n_\mathrm{l}$    & $n_\mathrm{r}$    & $h_\mathrm{0,l}$  & $h_\mathrm{0,2}$  & $k_\mathrm{sat,l}$  & $k_\mathrm{sat,r}$ & $\Delta \theta_\mathrm{l}$ & $\Delta \theta_\mathrm{r}$ & $\psi_\mathrm{l}$  & $\psi_\mathrm{r}$  & OF    \\
       & [$\mathrm{sm^{-1/3}}$]  & [$\mathrm{sm^{-1/3}}$]  & [$\mathrm{mm}$] & [$\mathrm{mm}$] & [$\mathrm{mmh^{-1}}$]   & [$\mathrm{mmh^{-1}}$]  & [-]   & [-]    & [mm]  & [mm]  & [-]  \\
1-2-3    & 0.036 & 0.019 & 9.17 & 5.25 & 8.986  & 3.44  & 0.65    & 0.04    & 92.34 & 8.33  & -0.99 \\
1-2     & 0.047 & 0.015 & 9.73 & 7.68 & 10.059 & 1.30  & 0.52    & 0.11    & 14.49 & 13.80 & -0.99 \\
1-3     & 0.068 & 0.015 & 5.36 & 7.68 & 5.395  & 1.78  & 0.64    & 0.11    & 15.26 & 13.80 & -0.98 \\
2-3     & 0.050 & 0.014 & 5.24 & 5.61 & 2.761  & 2.48  & 0.68    & 0.16    & 54.34 & 20.14 & -0.98 \\
1      & 0.040 & 0.013 & 9.95 & 5.51 & 3.191  & 3.12  & 0.77    & 0.27    & 0.61  & 0.17  & -0.99 \\
2      & 0.063 & 0.048 & 7.12 & 8.76 & 9.004  & 0.91  & 0.74    & 0.17    & 36.91 & 51.51 & -1.00 \\
3      & 0.032 & 0.020 & 6.44 & 5.14 & 6.630  & 3.36  & 0.73    & 0.04    & 38.00 & 23.21 & -0.97 
\end{tblr}
\caption{Near-optimal solutions for different combinations of gauges and for only 1 storm of 10.8~$\mathrm{mmh^{-1}}$ during 90 minutes. The known parameters are $n_l = 0.06~\mathrm{sm^{-1/3}}$, $n_r = 0.015$, $h_{0,l} = 1~\mathrm{mm}$, $h_{0,r} = 4~\mathrm{mm}$, $k_{\mathrm{sat,l}} = 8~\mathrm{mmh^{-1}}$, $k_{\mathrm{sat,r}} = 2~\mathrm{mmh^{-1}}$, $\Delta \theta_{\mathrm{l}} = 0.6$, $\Delta \theta_{\mathrm{r}} = 0.15$, $ \psi_{\mathrm{l}} = 20~\mathrm{mm}$, and  $ \psi_{\mathrm{r}} = 100~\mathrm{mm}.$}
\label{tab:only_one_storm}
\end{table*}

\begin{table*}
\small
\centering
\begin{tblr}{
  cell{1}{1} = {r=2}{},
  cell{1}{2} = {r=2}{},
  cell{1}{5} = {r=2}{},
  cell{3}{1} = {r=3}{},
  cell{3}{2} = {r=3}{},
  cell{3}{3} = {r=3}{},
  cell{3}{4} = {r=3}{},
  cell{6}{1} = {r=3}{},
  cell{6}{2} = {r=3}{},
  cell{6}{3} = {r=3}{},
  cell{6}{4} = {r=3}{},
  cell{9}{1} = {r=3}{},
  cell{9}{2} = {r=3}{},
  cell{9}{3} = {r=3}{},
  cell{9}{4} = {r=3}{},
  cell{12}{1} = {r=3}{},
  cell{12}{2} = {r=3}{},
  cell{12}{3} = {r=3}{},
  cell{12}{4} = {r=3}{},
  cell{15}{1} = {r=3}{},
  cell{15}{2} = {r=3}{},
  cell{15}{3} = {r=3}{},
  cell{15}{4} = {r=3}{},
  cell{18}{1} = {r=3}{},
  cell{18}{2} = {r=3}{},
  cell{18}{3} = {r=3}{},
  cell{18}{4} = {r=3}{},
  cell{21}{1} = {r=3}{},
  cell{21}{2} = {r=3}{},
  cell{21}{3} = {r=3}{},
  cell{21}{4} = {r=3}{},
  cell{24}{1} = {r=3}{},
  cell{24}{2} = {r=3}{},
  cell{24}{3} = {r=3}{},
  cell{24}{4} = {r=3}{},
  hline{1,3,6,9,12,15,18,21,24,27} = {-}{},
}
Event & {Rainfall Boundary\\~Condition} & {Rainfall \\ Duration} & {Rainfall\\Volume} & Gauge & NSE    & Vol Error & r\textsuperscript{2} & PBIAS  \\
      &                                 & {[}min]           & {[}mm]           &        & {[}-]  & {[}\%]    & {[}-]                & {[}\%] \\
1     & Steady State                    & 135               & 24.3             & Outlet & 0.999  & 1.076     & 1.000                & 0.347  \\
      &                                 &                   &                  & Left   & {[}-]  & {[}-]     & {[}-]                & {[}-]  \\
      &                                 &                   &                  & Right  & 0.999  & 0.951     & 0.999                & 0.589  \\
2     & Steady State                    & 45                & 8.1              & Outlet & 0.993  & 1.847     & 0.997                & 1.184  \\
      &                                 &                   &                  & Left   & {[}-]  & {[}-]     & {[}-]                & {[}-]  \\
      &                                 &                   &                  & Right  & 0.989  & 1.456     & 0.995                & 5.171  \\
3     & Steady State                    & 135               & 48.6             & Outlet & 0.998  & 2.121     & 1.000                & 0.494  \\
      &                                 &                   &                  & Left   & 0.990  & 5.961     & 1.000                & 1.269  \\
      &                                 &                   &                  & Right  & 0.999  & 0.224     & 1.000                & 0.381  \\
4     & Steady State                    & 45                & 36.45            & Outlet & 0.999  & 0.378     & 1.000                & 0.712  \\
      &                                 &                   &                  & Left   & 0.996  & 3.227     & 1.000                & 1.696  \\
      &                                 &                   &                  & Right  & 0.998  & -0.483    & 0.999                & 1.119  \\
5     & Huff 1st Quartile               & 135               & 24.3             & Outlet & 0.997  & -2.505    & 0.999                & 1.189  \\
      &                                 &                   &                  & Left   & 0.960  & -20.399   & 0.997                & 4.777  \\
      &                                 &                   &                  & Right  & 0.997  & 0.226     & 0.999                & 1.401  \\
6     & Huff 1st Quartile               & 45                & 8.1              & Outlet & 0.983  & -3.425    & 0.993                & 1.862  \\
      &                                 &                   &                  & Left   & -0.061 & {[}-]     & {[}-]                & 59.408 \\
      &                                 &                   &                  & Right  & 0.895  & 6.872     & 0.946                & 15.927 \\
7     & Huff 1st Quartile               & 135               & 48.6             & Outlet & 0.999  & 1.310     & 1.000                & 0.684  \\
      &                                 &                   &                  & Left   & 0.996  & 4.049     & 1.000                & 1.358  \\
      &                                 &                   &                  & Right  & 0.998  & 0.059     & 0.999                & 0.972  \\
8     & Huff 1st Quartile               & 45                & 36.45            & Outlet & 0.999  & 0.708     & 1.000                & 1.022  \\
      &                                 &                   &                  & Left   & 0.992  & 3.125     & 0.998                & 2.573  \\
      &                                 &                   &                  & Right  & 0.995  & -0.424    & 0.998                & 2.571  
\end{tblr}
\caption{Validation metrics for events outside of the calibration conditions. Each row represent results of one event, for different gauges. Events 1-2 are with the smallest rainfall intensity used in calibration (i.e., 10.8 $\mr{mmh^{-1}}$, while events 3-4 are with the largest one (i.e., 32.4~${\mr{mmh^{-1}}}$. Similarly, the events with unsteady-state rainfall follow the same pattern, resulting in the same rainfall volumes but temporally distributed with Huff 1st quartile hyetographs.}
\label{tab:validation}
\end{table*}

\subsection{Limitations, Challenges, and Opportunities of this Modeling Approach}
Calibrating a fully-distributed hydrodynamic and water quality model requires not only field observations, but also depends extensively on the quality and resolution of the terrain, soils, and land use and land cover models. However, as presented in this paper, this modeling approach can be easily applied worldwide in catchments with scarce time-series observations of rainfall and/or a representing variable of the flow dynamics such as discharge depths and/or a variable representing the pollutant such as the pollutant concentration. Since pollutant concentrations are inherently associated with accurate discharge modeling, calibrating water quantity and quality parameters altogether might result in high equifinality if only pollutographs are the optimization criteria.  

This approach can be enhanced and easily expanded by allowing calibration not only with time series but also with maps of flood extent, magnitude, or by socio-hydrological information \citep{fava2022linking} such as maximum depths in certain flood points, especially in catchments with no gauge stations \citep{gomes2023hydropol2d}. The challenge, however, is to find a suitable single objective cost function that can normalize different optimization criteria into a single and homogeneous cost function. In addition, the minimum requirement, however, is the rainfall intensity time-series in a proper resolution that depends on the catchment response. Regarding rainfall, this approach could also be improved by allowing space-variant rainfall that could be derived from radar, satellite imagery, or by interpolation of source-gauged rainfall stations.

It is recommended that a sensitivity analysis be performed before automatic calibration to avoid wasting computational resources on variables that do not play a substantial role in the hydrologic-hydrodynamic behavior of the catchment. Although the results presented in this paper indicate that some parameters might be more sensitive than others, the results can vary dramatically for different catchments with different topography and soil properties. 

The use of worldwide datasets to represent LULC and SOIL allows a proper definition of model parameters such as $n$ or $k_{\mathrm{sat}}$, as shown in \citep{soliman2022assessment,gupta2021global}. Studies such as the aforementioned ones might facilitate the parametrization of fully distributed models and can be opportunities for worldwide application.

\section{Conclusions}
An optimization-based algorithm was developed and applied to calibrate a fully distributed hydrological-hydrodynamic and water quality model (HydroPol2D). The algorithm can find near-optimal set of parameters to explain the observed gauged information, such as flow discharge, pollutant concentration, or flood depths. The results of Numerical Case Studies 1, 2, from the real-world case study in Numerical Case Study 3, and from the Equifinality analysis in Numerical Case Study 4 support the following:

\begin{itemize}
    \item A1: The model can accurately predict not only the Green-Ampt infiltration parameters but also Manning's roughness coefficients and initial abstraction values, as shown in Numerical Case Study 1. The Predicted hydrographs match, with $\mathrm{NSE > 0.99}$ the considered real observed hydrograph in Numerical Case Study 1.
    \item A2: The algorithm can find not only the wash-off parameters, but also the initial mass (error $< 15\%$) of the pollutant in the wooden-plane catchment to match the observed pollutgraph, as shown in Numerical Case Study 2. 
    \item A3: The model can still find a physically bounded near-optimal set of parameters to calibrate the observed hydrograph with NSE = 0.89, indicating good accuracy. In addition, using this set of parameters were possible to determine (i) infiltration maps, maximum flow velocities (ii), maximum flood depths, (iii) outlet hydrograph.
    \item A4-1: The equifinality problem is reduced by the addition of runoff-producing events and by choosing at least one gauge in hydrological unit regions. Using only the outlet gauge as the information might produce a feasible (i.e., within the parameter range) but wrong parameter set that explains the observed data. This set of parameters tends to produce small errors in peaks and hydrograph shapes and relatively larger errors in runoff volumes. The use of runoff-producing events with different flow parameters typically produces better parameter estimation. The parameter estimation error is reduced by a more reduced parameter range that can be attached by experts or by GIS available worldwide datasets to reduce parameter ranges.
    \item A4-2: The model presented accurate results when calibrated only with the outlet gauge hydrograph as the sole information for calibration. Although the equifinality is observed by the compensation of the infiltration parameters, the model presented acceptable results in most cases of different rainfall volumes, intensities, and distributions. A reduced model performance is observed for events with little or no observed runoff in the gauges. However, in general, the model presented accepted results in gauges not used for calibration for different rainfall durations, volumes, intensities, and rainfall distributions even with the model calibrated with only the outlet gauge. This indicates an opportunity to move towards conceptual and simplified lumped models flood assessment to physically-based, fully-distributed analysis, since both models can be calibrated with the same input data.
\end{itemize}

Therefore, the methods applied in this paper can be replicated in all catchments with observations at least at one point. The more points with runoff observations, the better the reduction of equifinality. Using only freely available datasets, this method can be applied for catchments with observations at gauging stations to extrapolate results in the whole catchment domain moving from typically limited lumped-parameter models to fully-distributed physically-based analysis. However, the methodology strategy developed in this paper is only applicable if some constraints are satisfied, such as:

\begin{itemize}
    \item The overland flow is predominantly hortonian.
    \item The effect of human made drainage systems such as reservoirs, dams, polders, or any other hydraulic structure operation does not govern the whole catchment hydrodynamics.
    \item The catchment can be modeled with space-invariant precipitation.
    \item The optimization cost function is relatively fast, allowing multiple evaluations in a reasonable time. 
\end{itemize}

The aforementioned requirements are typically satisfied in relatively small to mid-size catchments. Advancing these limitations and developing a framework capable of adapting to whatever data is available could help modelers use distributed models and improve flood and water quality spatial analysis. 

\section{Data Availability Statement}
Some or all data, models, or code generated or used during the
study are available in a repository or online in accordance with
funder data retention policies. All software, figures, and data can be freely downloaded in \citep{gomes2023hydropol2d}.

\section*{Acknowledgment}
The authors appreciate the support of the City of San Antonio, by the San Antonio River Authority, CAPES Ph.D Scholarship, and the PPGSHS PROEX Graduate Program.

\section*{Supplemental Materials}
Supplementary data related to this article can be found at \href{https://drive.google.com/file/d/1HX75uchSHT-Vv9O0axFxWEawxamC-k16/view?usp=sharing}{https://drive.google.com/file/d/1HX75uchSHT-Vv9O0axFxWEawxamC-k16/view?usp=sharing}. 

\bibliographystyle{elsarticle-num-names} 
\bibliography{cas-refs}





\end{document}